\documentclass[a4paper,11pt]{article}
\usepackage{jheppub}
\usepackage[utf8]{inputenc}
\usepackage{amsmath}
\usepackage{graphicx}
\usepackage{amssymb}
\usepackage{multirow}
\usepackage[normalem]{ulem}

\DeclareMathOperator{\diag}{diag}

\newcommand{\orcid}[1]{\href{https://orcid.org/#1}{#1}}

\newcommand{\e}[1]{\times10^{#1}}

\title{Solar Neutrinos and the Strongest Oscillation Constraints on Scalar NSI}

\author[1]{Peter B.~Denton\note{\orcid{0000-0002-5209-872X}},}
\author[2]{Alessio Giarnetti\note{\orcid{0000-0001-8487-8045}},}
\author[2]{and Davide Meloni\note{\orcid{0000-0001-7680-6957}}}

\affiliation[1]{High Energy Theory Group, Physics Department, Brookhaven National Laboratory, Upton, NY 11973, USA}
\affiliation[2]{Dipartimento di Matematica e Fisica, Universit\`a di Roma Tre Via della Vasca Navale 84, 00146 Rome, Italy}

\emailAdd{pdenton@bnl.gov}
\emailAdd{alessio.giarnetti@uniroma3.it}
\emailAdd{davide.meloni@uniroma3.it}

\makeatletter
\hypersetup{colorlinks=true,allcolors=[rgb]{1,0.56,0},pdftitle=\@title}
\makeatother

\abstract{
Scalar non-standard neutrino interactions (sNSI) is a scenario where neutrinos can develop a medium dependent contribution to their mass due to a new scalar mediator.
This scenario differs from the commonly discussed vector mediator case in that the oscillation effect scales with density rather than density and neutrino energy.
Thus the strongest oscillation constraint comes from solar neutrinos which experience the largest density in a neutrino oscillation experiment.
We derive constraints on all the sNSI parameters as well as the absolute neutrino mass scale by combining solar and reactor data and find solar neutrinos to be $>1$ order of magnitude more sensitive to sNSI than terrestrial probes such as long-baseline experiments.}

\begin{document}

\maketitle

\section{Introduction}
The three-flavor neutrino oscillation picture is starting to take shape and three of the four remaining unknowns -- the atmospheric mass ordering, the octant of $\theta_{23}$, and the value of $\delta$ -- are expected to be determined in the coming generation of oscillation experiments \cite{Denton:2022een}.
The fourth unknown parameter is the absolute mass scale of neutrinos and is only somewhat constrained by oscillation data and cosmological measurements \cite{Planck:2018vyg,DESI:2024mwx,Craig:2024tky,Denton:2023hkx}.
This makes it an important time to carefully examine the data to see if it is converging on the three-flavor picture or if new physics hints are emerging.

Many new physics scenarios are considered in the literature that affect neutrino oscillations in different ways.
While these may not represent the entire model space, they do have quite different scaling behaviors, hopefully providing fairly complete coverage of conceivable scenarios.
Popular scenarios are sterile neutrinos, which may lead to new oscillation frequencies that can be directly probed or can be indirectly probed by determining if the $3\times3$ part of the mixing matrix that can be measured is unitary or not.
If there are not new frequencies detected in neutrino mixing and the matrix seems to be unitary, there could also be new interactions of neutrinos with matter particles: up quarks, down quarks, and electrons.
If the new interaction is charged-current, strong constraints exist making it unlikely that neutrino oscillation experiments will provide the most sensitive probes, but neutral-current interactions are often best probed via neutrino oscillations or scatterings.
These are called non-standard neutrino interactions (NSI) \cite{Wolfenstein:1977ue}, for an overview of NSI see e.g.~\cite{Dev:2019anc}.

The majority of NSI studies have focused on NSI with a vector mediator (vNSI), first proposed in Wolfenstein's 1977 paper.
It also has the same Lorentz structure as the weak interaction and thus has the possibility to interfere with the regular matter effect \cite{Wolfenstein:1977ue}.
It has been pointed out that this leads to an important degeneracy regarding the determination of the mass ordering \cite{Bakhti:2014pva}, see also \cite{deGouvea:2000pqg,Miranda:2004nb,Coloma:2016gei,Coloma:2017egw,Coloma:2017ncl,Denton:2018xmq,Denton:2021vtf,Chaves:2021pey,Denton:2022nol}.

While NSI's with different Lorentz structures have been examined in scattering for some time, only recently has there become a growing interesting in scalar NSI (sNSI) in neutrino oscillation environments \cite{Ge:2018uhz,Babu:2019iml,Smirnov:2019cae,Arguelles:2019xgp,Suliga:2020lir,Venzor:2020ova,Medhi:2021wxj,Medhi:2022qmu,Sarkar:2022ujy,Dutta:2022fdt,Denton:2022pxt,Medhi:2023ebi,Singha:2023set,Sarker:2023qzp,Sarker:2024ytu,Dutta:2024hqq,ESSnuSB:2023lbg,Gupta:2023wct,Liao:2015rma}.
Ref.~\cite{Suliga:2020lir} investigated new scalar (and vector) interactions in the Sun, but not in the context of oscillations.
Notably \cite{Babu:2019iml} investigates the interplay of oscillation and non-oscillation constraints on sNSI under a variety of assumptions about the mediator mass, the nature of neutrinos, and whether the interaction is with electrons or nucleons.
There are key regions of parameter space where constraints from oscillations, particularly solar neutrinos, dominate.
The solar constraints were derived in \cite{Ge:2018uhz} by looking only at solar neutrino data from Borexino \cite{Borexino:2017rsf}.
Additional solar neutrino data sets exist, however, some of which are considerably more precise.
In addition, nearly all existing studies assumed the lightest neutrino mass, $m_1$ in the normal ordering (NO) and $m_3$ in the inverted ordering (IO) to be exactly massless\footnote{Ref.~\cite{Medhi:2023ebi} investigated the impact of neutrino masses in sNSI at future long-baseline accelerator neutrino experiments.}.

While the strength of the effect of vNSI tends to grow with neutrino energy and density of the background field, implying that accelerator neutrinos and atmospheric neutrinos are some of the best places to look for them, the effect of sNSI increases only with the density of the background field, thus we find that solar neutrinos are the best oscillation means of probing them.
In this paper, we perform the most complete oscillation study of sNSI to date by looking at solar neutrino data from pp experiments, Borexino, and SNO and use terrestrial information from KamLAND and Daya Bay to provide the most relevant constraints on sNSI.
We also investigate the role of the absolute mass scale on such constraints.
Finally, we discuss how these constraints may evolve in the future with additional terrestrial measurements of the standard oscillation parameters with JUNO.

\section{Scalar NSI}
Neutrino oscillations can be affected by the presence of new physics.
In particular, it is well known that the existence of NSIs can modify the oscillation probabilities when neutrinos travel through matter with sufficient density.
If such interactions are vectorial, their presence can modify the standard matter potential which affects the oscillations. 
However, it has recently attracted much attention the case in which the NSI have a different Lorentz structure, of scalar type. In this case, the effective Lagrangian reads:
\begin{equation}
{\cal L}^{eff}_{\rm scalar\ NSI} = \frac{y_f y_{\alpha\beta}}{m_\phi^2}(\bar{\nu}_\alpha  \nu_\beta)
( \bar {f}  f ) \,,
\label{eq:efflag_sc}
\end{equation}
where the $y$'s are the Yukawa couplings to matter fermions and neutrinos and $m_\phi$ is the mass of the scalar mediator of the UV complete model.
Here we are focused on mediator masses that are not too small, so long as $1/m_\phi$ is small compared to the oscillation environment, see e.g.~\cite{Babu:2019iml}\footnote{In the event that the mediator is very light, new unusual phenomenon could arise, see e.g.~\cite{Smirnov:2019cae,Venzor:2020ova,Davoudiasl:2023uiq}.}.
The presence of such a term modifies the neutrino Dirac equation in the following way:
\begin{equation}
\label{eq:dirac}
\bar{\nu}_\beta\left[i\partial_{\mu}\gamma^\mu+\left(M_{\beta\alpha}+\frac{\sum_f N_f y_f y_{\alpha\beta}}{m_\phi^2}\right)\right]\nu_\alpha=0\,,
\end{equation}
where $M_{\beta\alpha}$ is the Dirac mass matrix of the neutrinos and $N_f$ is the number density of fermion $f$. It is clear that, in the neutrino oscillation context, the neutrino mass matrix which appears in the time evolution equation turns out to be modified by scalar NSI. Thus, the Hamiltonian governing neutrino oscillations is modified from the diagonal $M^2$ term to $(M+\delta M)(M+\delta M)^\dagger$, where $\delta M$ encodes the effects of the scalar NSI.
In the flavor basis, this can be parameterized in the following way:
\begin{equation}
\label{eq:eta}
\delta M=\sqrt{\Delta m_{21,{\rm KL}}^2}
\begin{pmatrix}
\eta_{ee} & \eta_{e\mu} & \eta_{e\tau} \\
\eta_{e\mu}^* & \eta_{\mu\mu} & \eta_{\mu\tau} \\
\eta_{e\tau}^* & \eta_{\mu\tau}^* & \eta_{\tau\tau}
\end{pmatrix}\,,
\end{equation}
where we have normalized $\delta M$ with a factor $\sqrt{\Delta m^2_{21,{\rm KL}}}=\sqrt{7.54\e{-5}{\rm\ ev}^2}=8.7\e{-3}$ eV \cite{KamLAND:2013rgu}, in order to compare the effect of the NSI to the effect of the standard solar oscillation and to make the parameters of the model, $\eta_{\alpha\beta}$, dimensionless\footnote{Note that, if normalizing the sNSI matrix, it is important to normalize it to a fixed quantity, rather than the true value of a $\Delta m^2$ which will float during a study making the interpretation of the numerical results extremely challenging.}.
We have chosen the $\delta M$ matrix to be Hermitian, even though in principle it is a general complex matrix.
In particular the matrix is required to be Hermitian only if the UV complete mediator is a real scalar.
With our assumptions, the diagonal effective parameters $\eta_{\alpha\alpha}$ need to be real, while the off-diagonal ones are complex and can be written as $\eta_{\alpha\beta}=|\eta_{\alpha\beta}|e^{i \phi_{\alpha\beta}}$. It is important to point out that, when comparing eqs.~\ref{eq:dirac} and \ref{eq:eta}, the following matching holds:
\begin{equation}
\eta_{\alpha\beta}=\frac{y_{\alpha\beta}}{m_\phi^2\sqrt{|\Delta m^2_{21,{\rm KL}}|}}\sum_fN_fy_f\,.
\end{equation}
Here, it is clear that the $\eta_{\alpha\beta}$ parameters that we wish to constrain are proportional to the matter density $N_f$.

We now show how scalar NSI affect the solar neutrino oscillations.
We can start from the neutrino Hamiltonian in the flavor basis
\begin{align} 
H
 =\frac{1}{2E} \left[ \Tilde{M}^2 +
                  a \left( \begin{array}{ccc}
            1     & 0& 0 \\
            0  & 0 & 0 \\
            0 & 0 & 0 
                   \end{array} 
                   \right) \right]\,,
\label{eq:matter}
\end{align}
where
\begin{equation}
a=2\sqrt{2}G_F N_e E\,,
\label{eq:a}
\end{equation}
$G_F$ is Fermi's constant, and $N_e$ is the electron number density, while $\Tilde{M}$ is defined as
\begin{equation}
    \Tilde{M}=U M  U^\dagger +\delta M\,,
\end{equation}
with $U$ being the PMNS mixing matrix and $M=\diag(m_1,m_2,m_3)$ the diagonal neutrino mass matrix. Given the definition of $\Tilde{M}$ it is clear that in this framework we can no longer subtract a diagonal matrix proportional to $m_1$ in order to write all the oscillation probabilities only in terms of mass splittings. This means that in presence of scalar NSI, neutrino oscillations depends also on the absolute neutrino mass scale. 

\subsection{Scalar NSI in the Sun}
We want now to understand what is the effects of the scalar NSI on the solar neutrino survival probabilities.
Since, as already pointed out, the scalar NSI parameters depends on the matter density, neutrino produced at different radii in the Sun are affected differently by such interactions.
We assume that $\phi$ couples equally to up and down quarks and not to electrons, thus the $\sum_fN_fy_f$ sum is proportional to the density.
A different flavor structure for the couplings to matter fermions will change the picture somewhat because the neutron fraction evolves with radius in the Sun and different neutrino populations are produced dominantly at different radii, although this effect is not large.
Thus, we further parametrize the new physics parameters in the following way: 
\begin{equation}
    \eta_{\alpha\beta}(r)=\frac{\rho(r)}{\rho^\odot}\eta_{\alpha\beta}^\odot\,,
\label{eq:rescaling}
\end{equation}
where $\rho^\odot$ is a typical solar density taken to be 100 g/cc for convenience and $\eta_{\alpha\beta}^\odot$ is now the parameters which encodes the magnitude of the sNSI.
Thus to relate $\eta^\odot$ to the fundamental parameters, we use
\begin{equation}
\frac{y_qy_{\alpha\beta}}{m_\phi^2}=6.7\e{-15}{\rm\ eV}^{-2}\eta_{\alpha\beta}^\odot\,,
\end{equation}
where the prefactor rescales to the appropriate units.
We can also relate this relevant value in the Sun to that in the Earth used in \cite{Denton:2022pxt} by 
\begin{equation}
\eta_{\alpha\beta}^\oplus=5.2\e{-3}\eta_{\alpha\beta}^\odot\,,
\end{equation}
where $\eta^\oplus$ is defined the same way as above but using $\Delta m^2_{31}$ instead of $\Delta m^2_{21}$ in eq.~\ref{eq:eta} and $\rho^\oplus=3$ g/cc instead of $\rho^\odot=100$ g/cc.

\begin{figure}
    \centering
    \includegraphics[width=15.5cm, height=6cm]{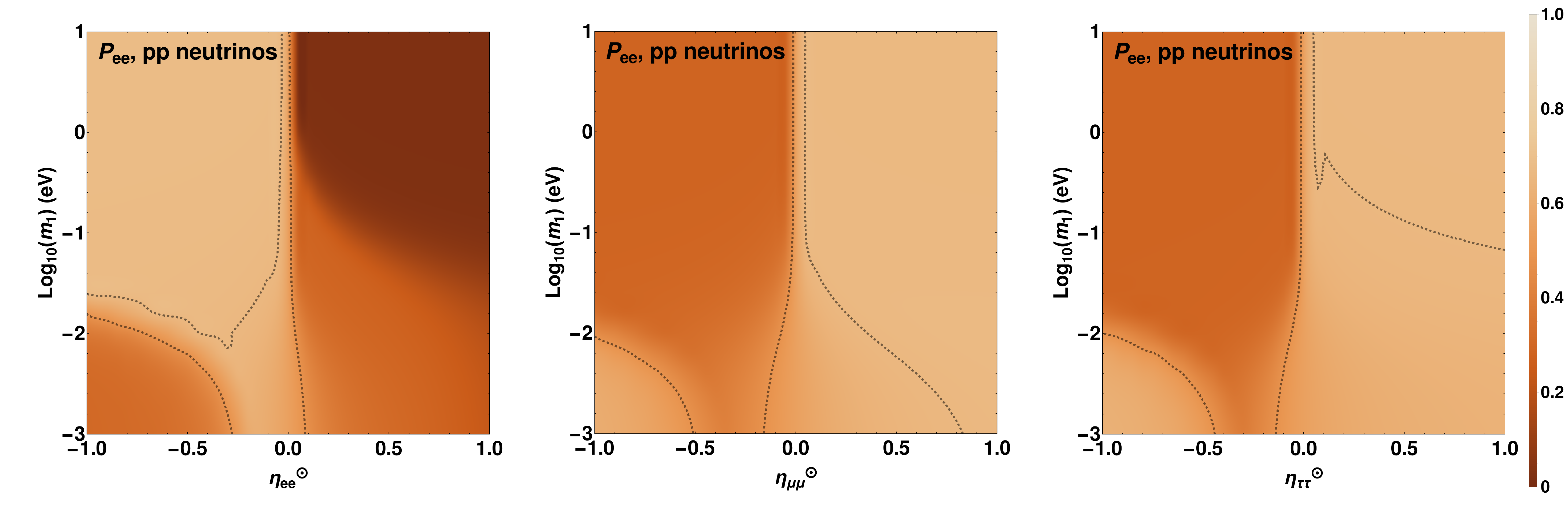}
    \includegraphics[width=15.5cm, height=6cm]{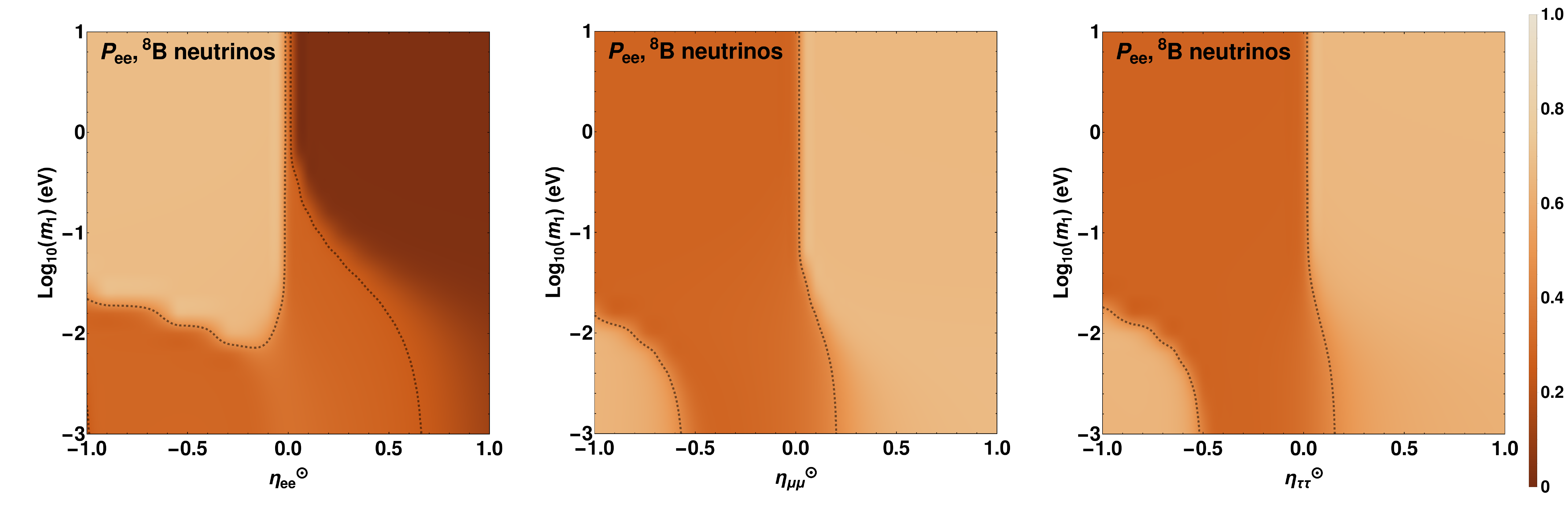}
     \caption{Survival probability for solar neutrinos as a function of the lightest neutrino mass $m_1$ and $\eta_{ee}^\odot$ (left), $\eta_{\mu\mu}^\odot$ (middle) and $\eta_{\tau\tau}^\odot$ (right). In the top panels we show the probability for pp neutrinos while in the bottom panels the probability for $^8$B neutrinos. The dashed curves represents the region of the parameter space allowed by current measurements.}
    \label{fig:diagonal_prob}
\end{figure}

\begin{figure}
    \centering
    \includegraphics[width=15.5cm, height=6cm]{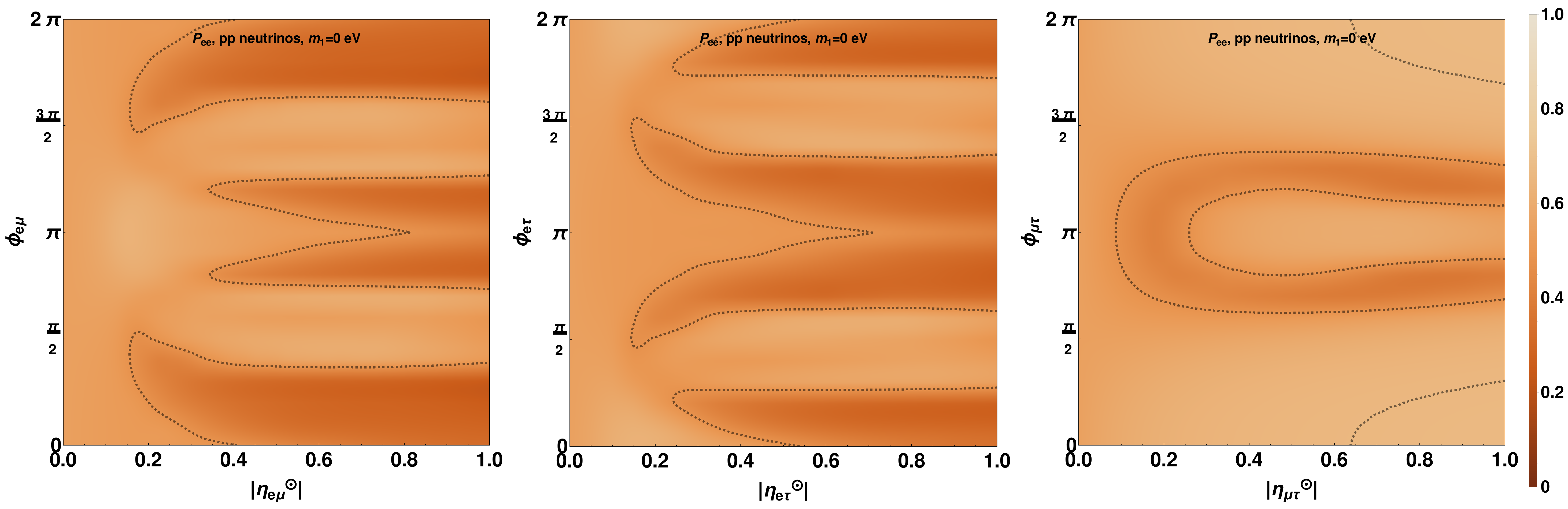}
    \includegraphics[width=15.5cm, height=6cm]{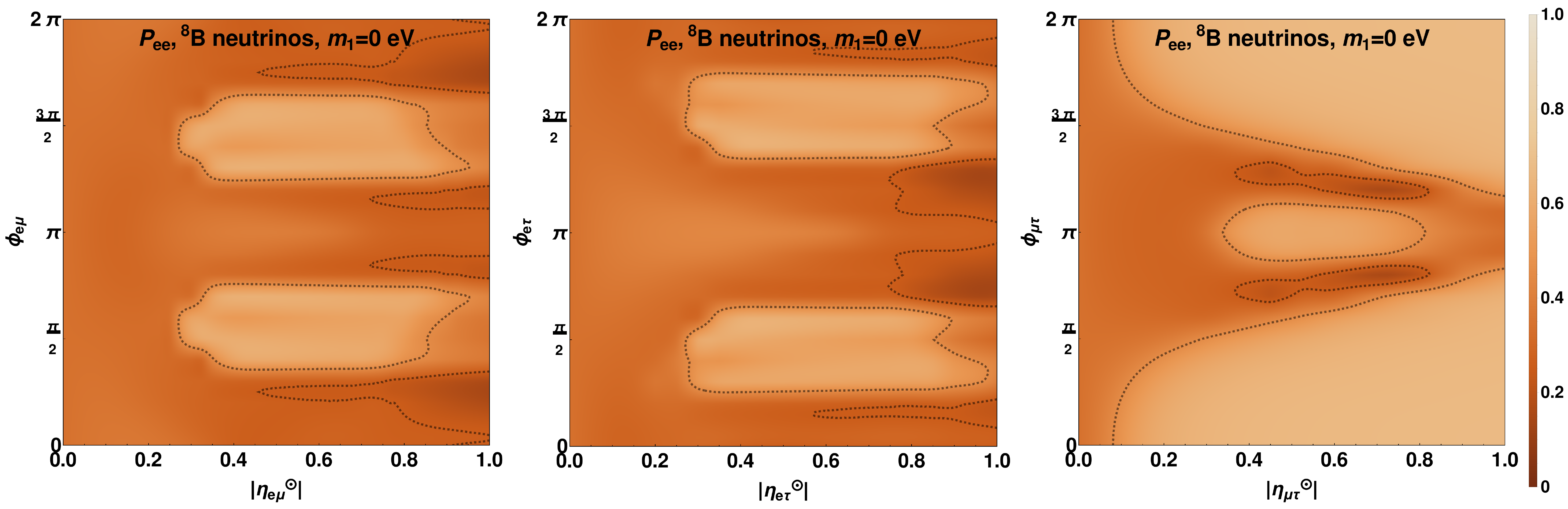}
    \caption{Survival probability for solar neutrinos as a  function of $\eta_{e\mu}^\odot$ (left) $\eta_{e\tau}^\odot$ (middle) and $\eta_{\mu\tau}^\odot$ (right) and their phases for $m_1=0$ eV. In the top panels we show the probability for pp neutrinos while in the bottom ones the probability for $^8$B neutrinos.}
    \label{fig:nondiagonal_prob}
\end{figure}

\begin{figure}
    \centering
    \includegraphics[width=15.5cm, height=6cm]{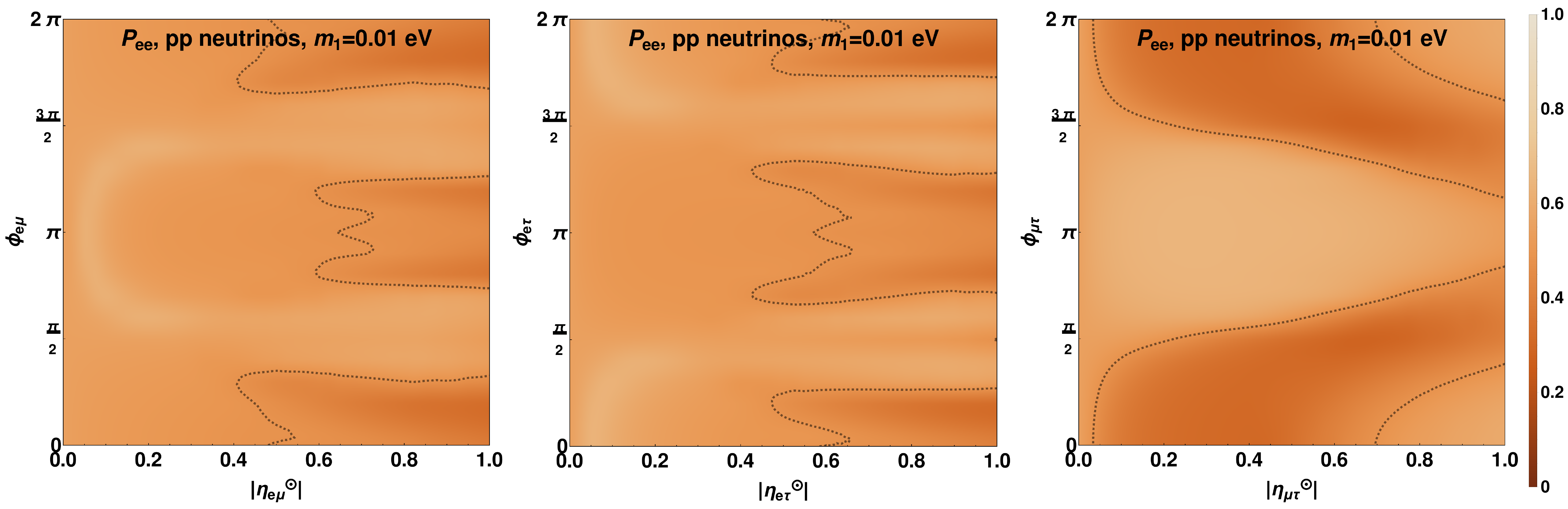}
    \includegraphics[width=15.5cm, height=6cm]{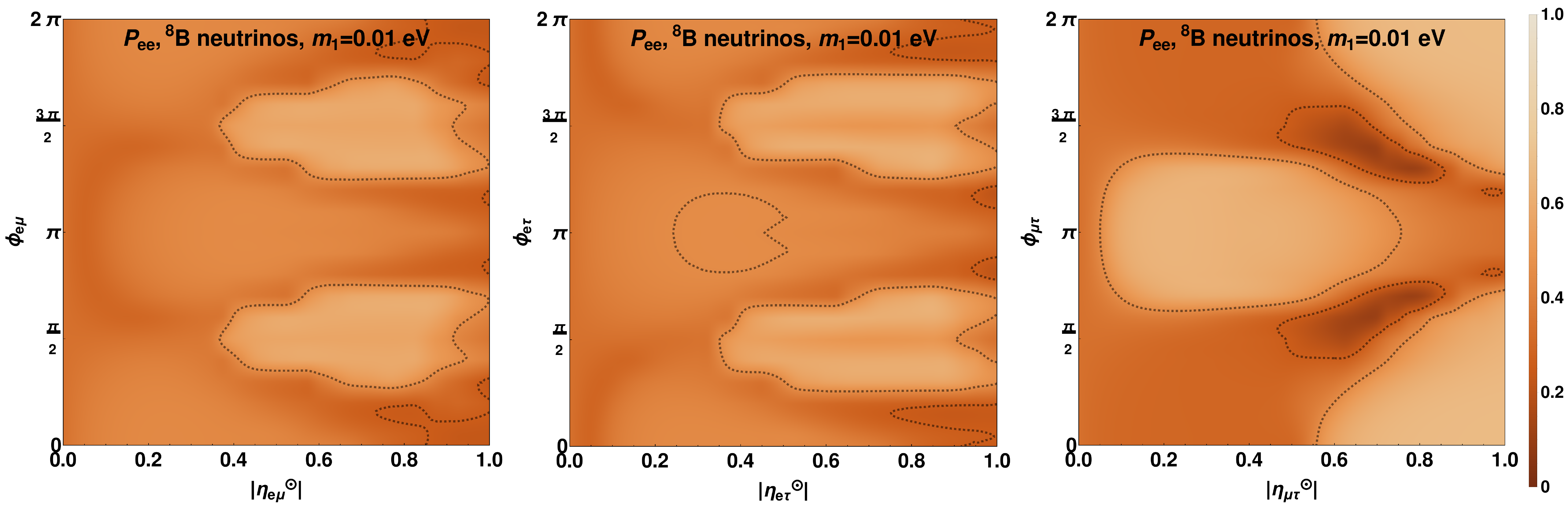}
    \caption{Same as Fig.~\ref{fig:nondiagonal_prob} but for $m_1=0.01$ eV. }
    \label{fig:nondiagonal_prob_mass}
\end{figure}

\section{Solar neutrinos}

Solar neutrinos, which begin as electron neutrinos, are produced in the inner dense part of the Sun from various different nuclear processes.
We focus on neutrinos coming from the four different channels which provide the most information about neutrino properties:
\begin{itemize}
    \item \textbf{pp neutrinos}: these are the most abundant and are produced from the reaction $pp\to d e^+ \nu$. Since in the final state we have three particles, the neutrino energy is not fixed. However, for our purposes, we can fix the pp-neutrinos energy to their mean energy of 0.267 MeV;
    \item \textbf{$^7$Be neutrinos}: neutrinos produced by the Beryllium 7 electron capture. These neutrinos have a fixed energy of 0.826 MeV
    \item \textbf{pep neutrinos}: such neutrinos are produced in the reaction $p e^- p\to d\nu$ and their flux is two order of magnitude less than the pp case. For our purposes, the neutrino energy is fixed at 1.44 MeV
    \item \textbf{$^8$B neutrinos}: in this case neutrinos are produced by the $\beta$-decays of boron nuclei. Their flux is four order of magnitude smaller than the pp one but the mean neutrino energy is around 8.1 MeV.
\end{itemize}
These four kind of solar neutrinos are particularly interesting because they are affected in different ways by the matter potential inside the Sun due to their range of energies. Indeed, while for the less energetic pp neutrinos the Sun does not affect their oscillations in the conventional picture, the $^8$B neutrino undergo the MSW transition due to the presence of matter in the usual picture. On the other hand, the intermediate pep neutrinos have an energy which allow us to probe the energy region in which the transition occurs. In the adiabatic approximation, characterized by slowly changing matter density, it is possible to compute the oscillation disappearance probability for a terrestrial experiment. In the two flavor approximation, averaging out the oscillation occurring during the long distance between the Sun and the Earth and neglecting the so-called ``jump probability" \cite{Parke:1986jy,Haxton:1986dm,Petcov:1987xd}, we have:
\begin{equation}
    P_{ee}^{2\nu}=\frac{1}{2}(1+\cos{2\theta_{12}^s}\cos{2\theta_{12}})\,,
    \label{eq:Pee2nu}
\end{equation}
where $\theta^s$ is the effective mixing angle modified by the matter potential in the production point of the initial neutrino and $\theta_{12}$ is the relevant mixing angle in vacuum. At a given matter potential, the effective matter mixing angle reads:
\begin{equation}
    \cos{2\theta_{12}^s}=\frac{\cos{2\theta_{12}}-\Tilde{A}}{\sqrt{(\cos{2\theta_{12}}-\Tilde{A})^2+\sin^2{2\theta_{12}}}}\,,
    \label{eq:theta12s}
\end{equation}
in which we have defined:
\begin{equation}
    \Tilde{A}=\frac{a}{\Delta m^2_{21}}\,,
    \label{eq:tildeA}
\end{equation}
where $E_\nu$ is the neutrino energy and $\Delta m^2_{21}$ is the relevant mass-squared splitting between the neutrino masses. It is clear that if $\Tilde{A}\ll\cos{2\theta_{12}}$, the effective mixing angle coincides with the vacuum one. In this regime, the survival probability is simply:
\begin{equation}
    P_{ee}^{vac}=\frac{1}{2}(1+\cos^2{2\theta_{12}}) \,.
\end{equation}
It can be observed that for the pp neutrinos, which have an energy of around 0.1 MeV, the condition $\Tilde{A}\ll\cos{2\theta_{12}}$ holds for the matter density of the Sun's core ($\sim$100 $g/cm^3$).

In the opposite regime, where matter effects are strong enough to overcome the standard oscillation amplitude, namely when $\Tilde{A}\gg\cos{2\theta_{12}}$, we have that $\cos{2\theta_{12}^s}\sim-1$. Thus, the survival probability becomes:
\begin{equation}
    P_{ee}^{matt}=\sin^2{\theta_{12}}\,,
\end{equation}
that is what we observe for more energetic neutrinos like the $^8$B neutrinos.

This simple approach is very useful and can be used to explain the solar neutrino data \cite{Mikheyev:1985zog}. However, in the intermediate regime in which neither of the two extreme conditions, $\Tilde{A}\ll\cos{2\theta_{12}}$ nor $\Tilde{A}\gg\cos{2\theta_{12}}$, hold, it is important to take into account the production point of each neutrino in order to compute the survival probability. Indeed, for each point inside the Sun, $\cos{2\theta_{12}^s}$ is different. Therefore, we can introduce the quantity:
\begin{equation}
    \overline{\cos2\theta_{12}^s}=\int_{0}^{R} \phi(r) \cos2\theta_{12}^s(r)dr\,,
\end{equation}
where $R$ is the Sun radius and $\phi(r)$ is a normalized function describing where neutrinos are produced in the Sun.
The radial dependence in $\theta_{12}^s(r)$ comes from the fact that $\theta_{12}^s$ depends on the matter effect (see eq.~\ref{eq:theta12s}) which in turn depends on the electron density (see eqs.~\ref{eq:tildeA} and \ref{eq:a}), and the electron density depends on the radius of the Sun as given by the Standard Solar Model.
In our analysis we use the BS05(OP) model \cite{Bahcall:2004pz}.
The survival probability in eq.~\ref{eq:Pee2nu} is then slightly modified to:
\begin{equation}
    P_{ee}^{2\nu}=\frac{1}{2}(1+\overline{\cos{2\theta_{12}^s}}\cos{2\theta_{12}})\,.
\label{2nuprob}
\end{equation}

Let us now discuss the modifications of the probabilities in the full $3\nu$ framework. It can be shown that in the two regimes, vacuum and matter dominated, the effects of the third neutrino flavor can be taken into account using the formula: 
\begin{equation}
    P_{ee}=\cos^4\theta_{13}P_{ee}^{2\nu}+\sin^4\theta_{13}\,,
\label{2flavcorr}
\end{equation}
where $P_{ee}^{2\nu}$ is the two-flavor probability in eq.~\ref{2nuprob} and $\cos^2\theta_{13}\tilde A$ is the relevant three-flavor matter potential.
See e.g.~\cite{Lim:1987yd,Akhmedov:1999uz}.

\begin{figure}
    \centering
    \includegraphics[width=0.49\textwidth]{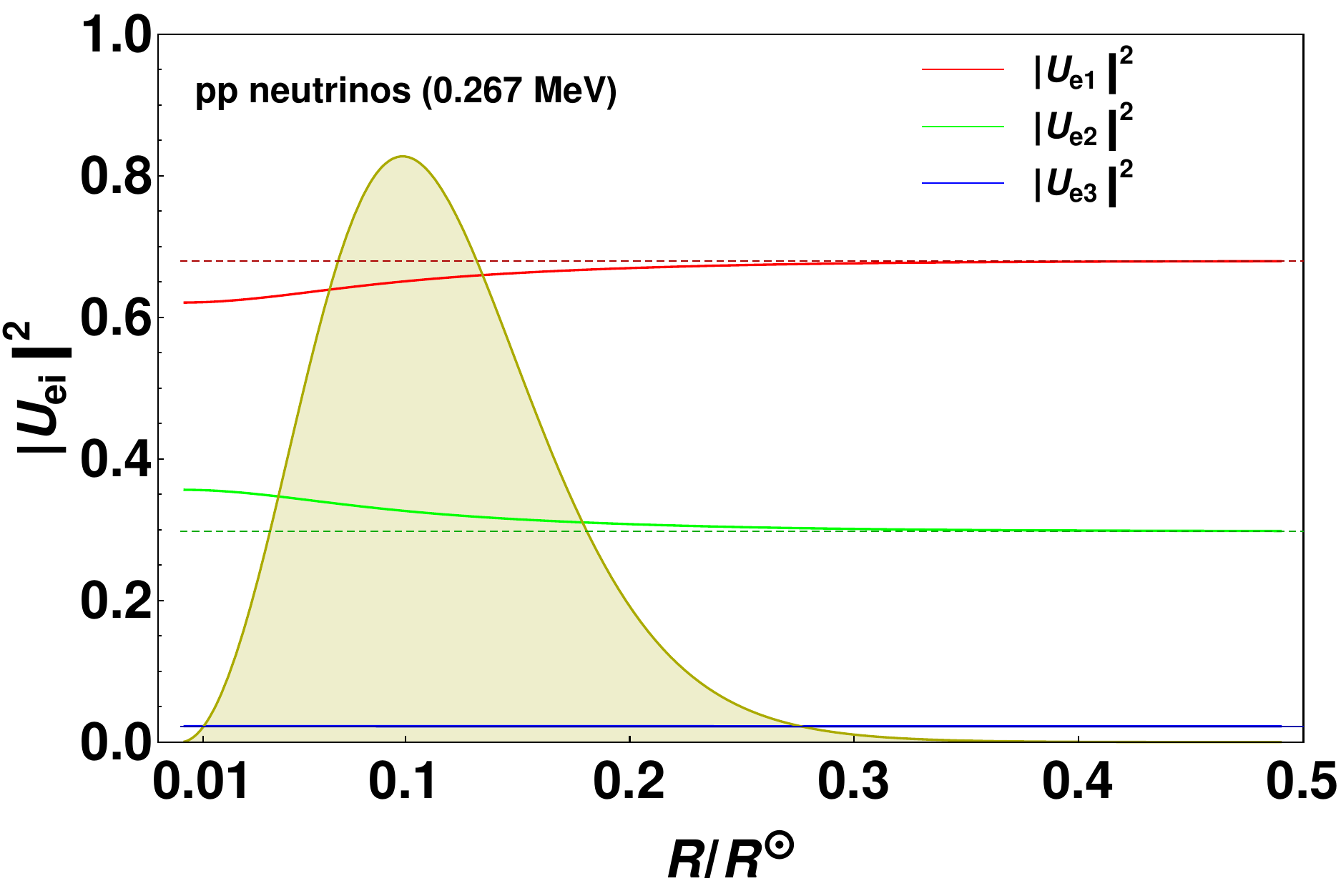}
     \includegraphics[width=0.49\textwidth]{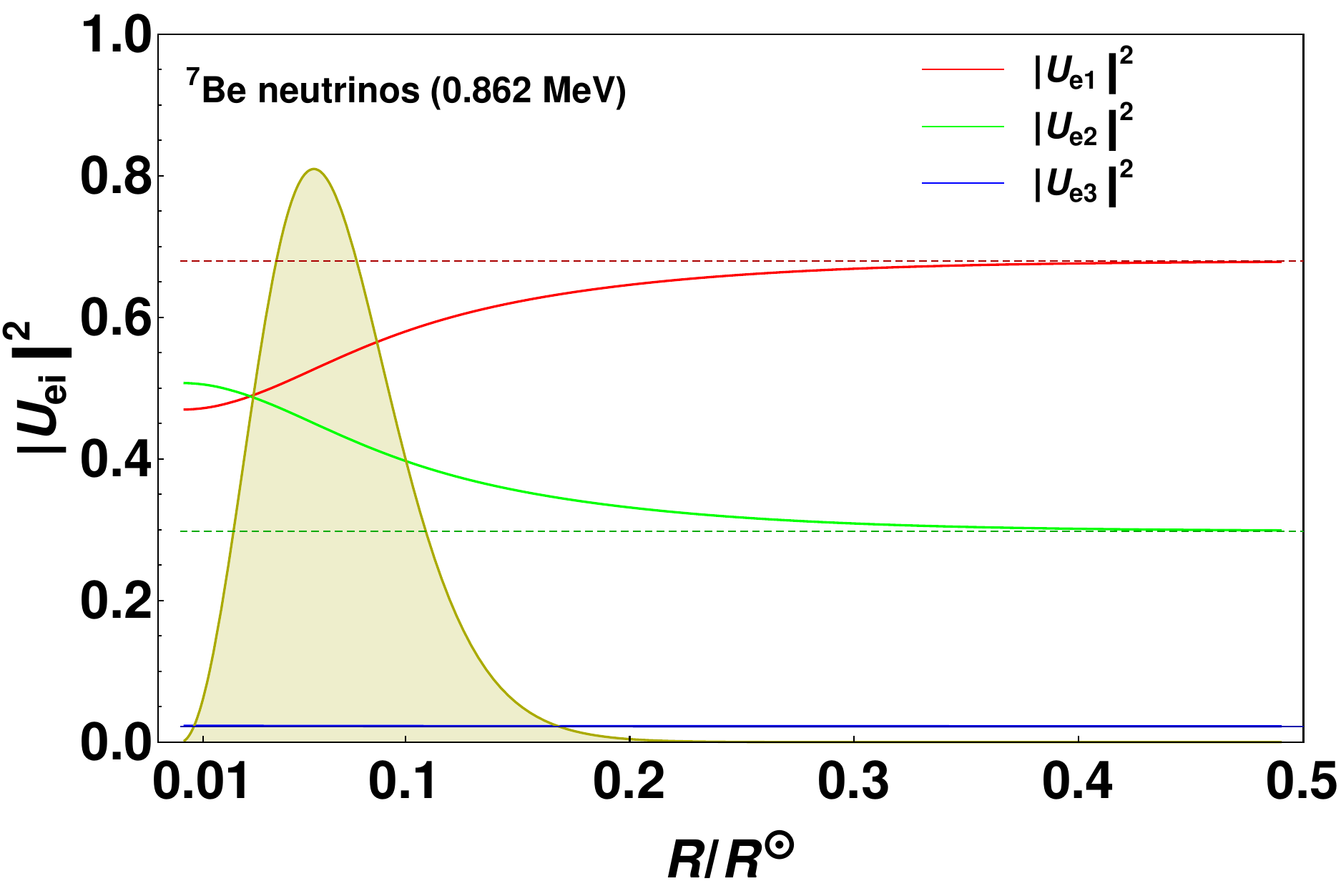}
    \includegraphics[width=0.49\textwidth]{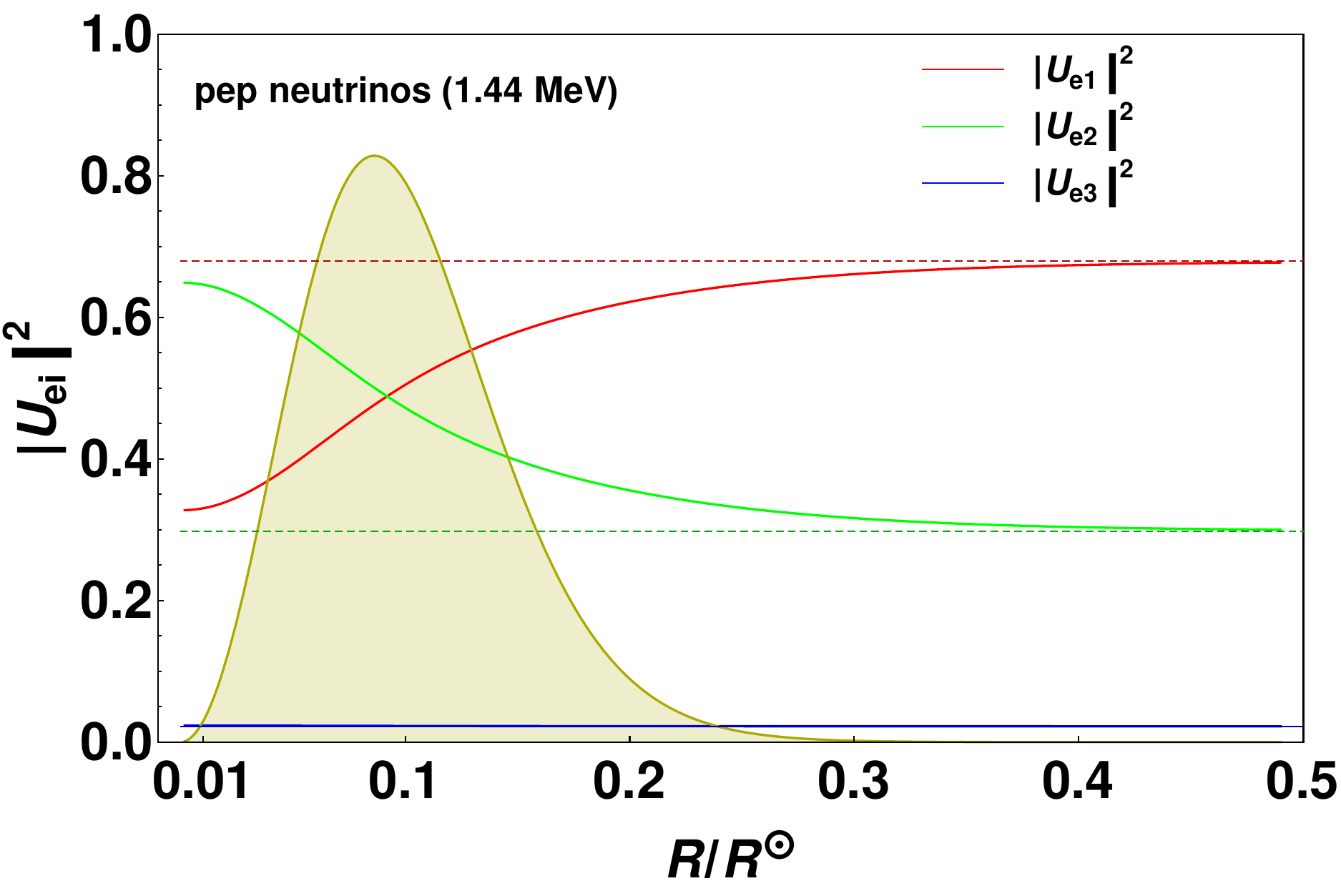}
    \includegraphics[width=0.49\textwidth]{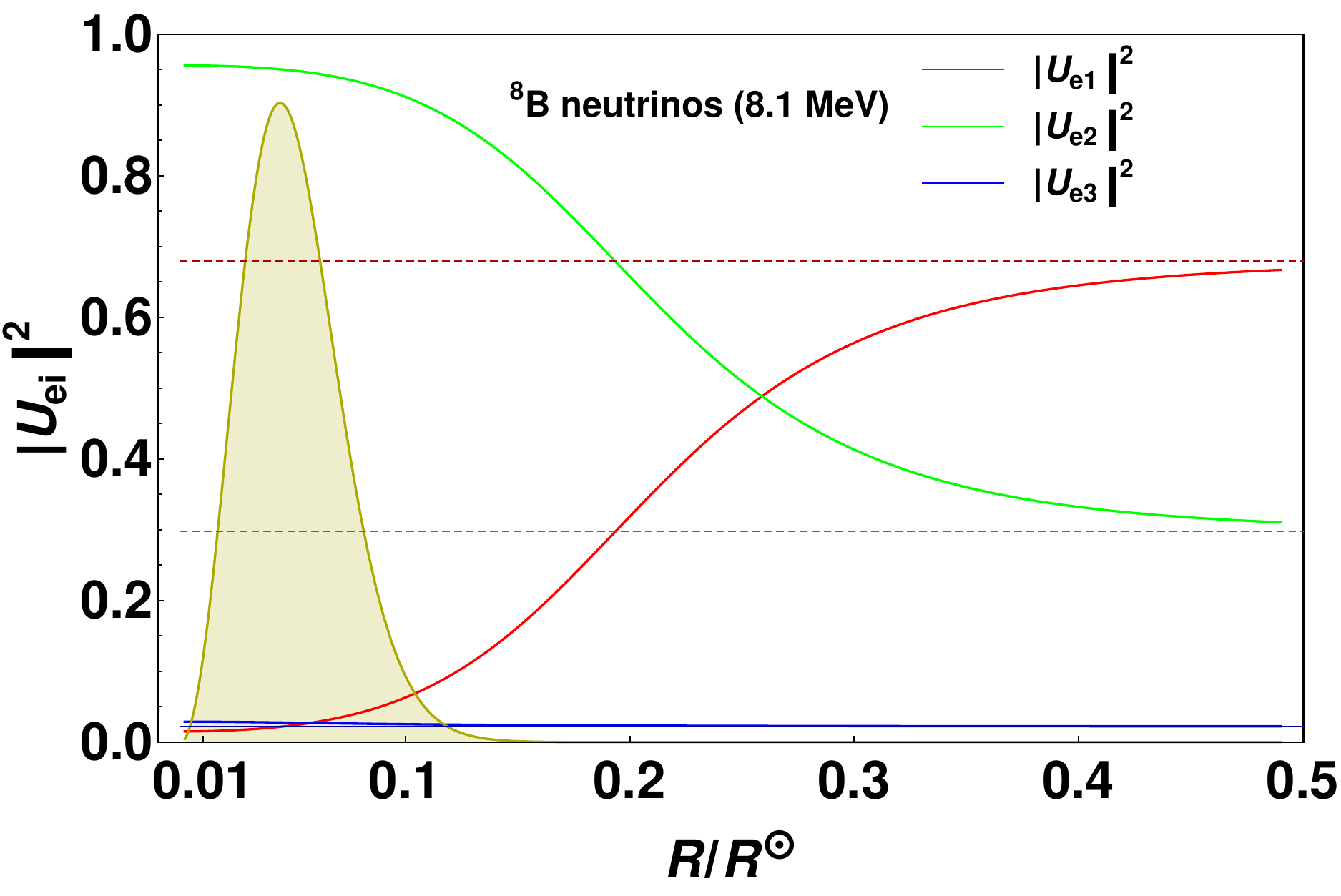}
    \caption{Values of the mixing matrix element $|U_{ei}|^2$ in matter as a function of the neutrino production point for different neutrino energies, corresponding to pp neutrinos (top left panel), $^7$Be neutrinos (top right panel), pep neutrinos (bottom left panel) and $^8$B neutrinos (bottom right panel). We also show with dashed lines the values of the mixing matrix elements in vacuum and the production point functions for the three neutrino components in arbitrary units (yellow region).}
    \label{matrixelem}
\end{figure}

This simplified formula, however, does not allow us to easily implement new physics. In order to take into account the 3-flavors effects in a more general framework, the oscillation probabilities can be computed in the following way:
\begin{equation}
    P_{ee}=\int_{0}^{R} \phi(r) \sum_{i=1}^3 |U_{ei}^s(r)|^2|U_{ei}|^2dr\,,
\label{3flav}
\end{equation}
where $U_{ei}^s$ are the mixing matrix elements in matter and $U_{ei}$ are those in vacuum. In table~\ref{Tabprob} we report the values of the survival probabilities for the four solar neutrino components computed using the 2-flavor approximation, the 2-flavor formula with the correction and the 3-flavor full probability. The values of the oscillation parameters are taken from table~\ref{oscpar}. In Fig.~\ref{matrixelem} we compute for a fixed neutrino energy corresponding to the four solar neutrino components, the value of the three mixing matrix elements $|U_{e1}|^2$, $|U_{e2}|^2$, and $|U_{e3}|^2$ as a  function of the neutrino production point inside the Sun. We also show with dashed lines the values of such matrix elements in vacuum and we superimpose the neutrino production functions $\phi(r)$ in arbitrary units. It is clear that for less energetic neutrinos (pp neutrinos), the matter potential mildly modifies the vacuum values of the mixing matrix elements. On the other hand, for pep neutrinos, at an energy of around 1.5 MeV, the transition between the two matrix elements $|U_{e1}|^2$ and $|U_{e2}|^2$ occurs at a radius which is around the production points of pep neutrinos. Finally, for the $^8$B neutrinos at a higher energy (8.1 MeV) the matter transition has completely occurred: $|U_{e1}|^2$ is close to one, while $|U_{e2}|^2$ is almost 0. It is worth to mention that, as expected, the 3-flavors correction on $|U_{e3}|^2$ remains $\sim2\%$, largely independent of the radius and density, as expected.

\begin{table}
\centering
\caption{\label{oscpar}Oscillation parameters used to compute the survival probabilities.}
\begin{tabular}{|c|c|c|c|c|}
\hline
\textbf{$\theta_{12}$ ($^\circ$)} & \textbf{$\theta_{13}$ ($^\circ$)} & \textbf{$\theta_{23}$ ($^\circ$)} & \textbf{$\Delta m_{21}^2$ ($eV^2$)} & \textbf{$\Delta m_{31}^2$ ($eV^2$)} \\ \hline
33.5                          & 8.6                           & 49.1                          & $7.49 \times 10^{-5}$               & $2.5 \times 10^{-3}$               \\ \hline
\end{tabular}
\end{table}

\begin{table}
\centering
\caption{\label{Tabprob}Expected survival probabilities computed for the three solar neutrino components using eq.~\ref{2nuprob} (left column), \ref{2flavcorr} (middle column) and \ref{3flav} (right column). Values for the oscillation parameters are taken from table \ref{oscpar}.}
\begin{tabular}{|lccc|}
\hline
\multicolumn{4}{|c|}{$P_{ee}$}                                                                                                              \\ \hline
\multicolumn{1}{|l|}{}       & \multicolumn{1}{l|}{2-flavors} & \multicolumn{1}{l|}{2-flavors+corrections} & \multicolumn{1}{l|}{3 flavors} \\ \hline
\multicolumn{1}{|l|}{pp}     & \multicolumn{1}{c|}{0.56}      & \multicolumn{1}{c|}{0.54}                  & 0.54 
 \\ \hline
\multicolumn{1}{|l|}{$^7$Be}    & \multicolumn{1}{c|}{0.51}      & \multicolumn{1}{c|}{0.50}                  & 0.50       \\ \hline
\multicolumn{1}{|l|}{pep}    & \multicolumn{1}{c|}{0.50}      & \multicolumn{1}{c|}{0.48}                  & 0.49                           \\ \hline
\multicolumn{1}{|l|}{$^8$B} & \multicolumn{1}{c|}{0.31}      & \multicolumn{1}{c|}{0.30}                  & 0.30                           \\ \hline
\end{tabular}
\end{table}

\subsection{Solar neutrino oscillation in presence of scalar NSI}
\label{sec:analytics}

In this section we will discuss the effect of scalar NSI on the solar neutrino oscillation. In general, the modification to the oscillation probabilities given by scalar NSI are very complicated to evaluate. Indeed, scalar NSI parameters can drastically alter the neutrino flavor transition; for instance, in presence of matter, a combination of $\eta_{\alpha\beta}$ might mimic the effect of the neutrino mass splittings and produce the observed oscillatory pattern even if the neutrino mass states are degenerate.
In Fig.~\ref{fig:diagonal_prob} we show the survival probabilities $P_{ee}$ in the $\eta_{\alpha\alpha}^\odot$ vs $m_1$ plane, where $m_1$ is the lightest neutrino mass (NO is assumed here). We show them for pp and $^8$B neutrinos.
Furthermore, in Figs.~\ref{fig:nondiagonal_prob} and \ref{fig:nondiagonal_prob_mass} we show the survival probabilities in the planes $\eta_{\alpha\beta}^\odot$ vs.~$\phi_{\alpha\beta}$ obtained for $m_1=0$ eV and $m_1=0.01$ eV, respectively. In all three figures, we show with dashed lines the parameters spaces for which we obtain survival probabilities compatible with the experimental ones \cite{BOREXINO:2018ohr}.

To understand the physics of sNSI in the Sun, we develop some approximations; the numerical work below is carried out using the full calculations.
Under some approximations, we can try to understand analytically the effect of the scalar NSI on the effective mass splitting and mixing angle for the solar neutrinos.
First, for sake of simplicity, in the following discussion we will take into account only two neutrino flavors.
The analytical results in this case are more readable and will allow to draw some conclusions on what to expect from our simulations.
The main disadvantage of this approach, however, is that not all the 6 scalar NSI parameters can be discussed.
The second approximation we adopt in this subsection is to expand our results in around $\eta=0$ and only take the leading correction to the observables.
This again does not allow to catch all the possible effects of scalar NSI, which in principle can be large.
However, in the context of solar neutrinos, we can expect that the least energetic \textit{pp}-neutrinos are not affected neither by usual matter effects nor by the scalar NSI with solar matter, in part due to the lower energies and in part because \textit{pp} neutrinos come from larger radii, and thus lower densities, than e.g.~$^8$B neutrinos.
Thus, large values of the $\eta$ parameters will not be too relevant for the measurement of \textit{pp}-neutrinos and our expansion is quite reliable in some useful regimes.
Finally, with the rescaling adopted in eq.~\ref{eq:rescaling}, the scalar NSI parameters could reach values greater than 1; this in principle breaks our series expansion.
Nonetheless, the choice of the rescaling density is arbitrary and, again, we can still draw our conclusions following our approach.

Given the above mentioned hypotheses, the neutrino mixing in this case will be described by a single mixing angle and a single mass splitting. Then, the mass matrix shift due to scalar NSI can be written as
\begin{equation}
 \delta M=\bar{m} 
    \begin{pmatrix}
       \eta_{1} & \eta_{3} \\
       \eta_{3}^\ast & \eta_{2}
    \end{pmatrix}
\end{equation}
where $\bar{m}$ is a normalization mass (for our analyses we used $\bar{m}=\sqrt{\Delta m_{21,{\rm KL}}^2}$, see eq.~\ref{eq:eta}) and the parameters $\eta_{1,2,3}$ are the scalar NSI parameters where we dropped the flavor indices to clearly distinguish them from the $\eta_{\alpha\beta}$ parameters in our full 3-flavors analyses.

Considering only the parameter $\eta_1$, we obtain the following correction to the effective mass splitting in matter
\begin{equation}
    [\Delta m^2_{21,\rm{eff}}]_{\eta_1}=2\bar{m}\, \eta_1\frac{(\Tilde{A}-1)m_1 \cos^2\theta_{12}+(\Tilde{A}+1) m_2 \sin^2\theta_{12}}{\sqrt{(\cos2\theta_{12}-\Tilde{A})^2+\sin^2 2\theta_{12}}}\,,
\end{equation}
where $\Tilde{A}$ has been defined in eq.~\ref{eq:tildeA} and $m_1$, $m_2$ are the two neutrino absolute masses (with $\Delta m^2=m_2^2-m_1^2$).
In the simple limit $m_1\to0$, this correction always has the same sign as $\eta_1$.
Since we do observe the matter transition in solar data, with a depletion of the oscillation probability for high-energy $^8$B neutrinos, the effective mass splitting cannot be too large.
This set a stringent upper limit on $\eta_1$.
However, if $\eta_1<0$, data can still be fitted with a larger $\Delta m^2$.
For this reason, the lower limit on $\eta_1$ set by solar data is less stringent than the upper one for $m_1\to0$.
In the case of a non-vanishing $m_1$, on the other hand, for high energetic solar neutrinos, $\Tilde{A}>1$ and the correction has the same sign as $\eta_1$, but it is larger, reducing the allowed values for the scalar NSI parameters.
Let us now consider the effective mixing angle in presence of $\eta_1$.
In this case we obtain the correction
\begin{equation}
    [\cos2\theta_{12}^s]_{\eta_1}=\bar{m}\, \eta_1 \, \sin^2 2\theta_{12} \frac{(\Tilde{A}+1)m_1 -(\Tilde{A}-1) m_2 }{\Delta m^2_{21} [(\cos2\theta_{12}-\Tilde{A})^2+\sin^2 2\theta_{12}]^{3/2}} \,.
    \label{eq:theffeta1}
\end{equation}
This correction, for $m_1\to0$, has again the opposite sign as $\eta_1$. Given that $\cos2\theta_{12}^s\to-1$ for $^8$B neutrinos, we can observe that the disappearance probability in eq.~\ref{2nuprob} can be fitted with smaller values of $\theta_{12}$ if $\eta_1>0$. Larger values of the mixing angle cannot easily accommodate the data since they would significantly increase $\Delta m^2_{21,\rm{eff}}$ pushing the matter transition towards higher energies. An increase of $m_1$ also increases the value of the effective mixing angle.

Let us now consider the case in which $\eta_2\neq0$.
We obtain the following corrections for the mass splitting and the mixing angle
\begin{equation}
\begin{split}
    [\Delta m^2_{21,\rm{eff}}]_{\eta_2}&=2\bar{m}\, \eta_2\frac{(\Tilde{A}-1)m_2 \cos^2\theta_{12}+(\Tilde{A}+1) m_1 \sin^2\theta_{12}}{\sqrt{(\cos2\theta_{12}-\Tilde{A})^2+\sin^2 2\theta_{12}}} \\
    [\cos2\theta_{12}^s]_{\eta_2}&=\bar{m}\, \eta_2 \, \sin^2 2\theta_{12} \frac{(\Tilde{A}+1)m_2 -(\Tilde{A}-1) m_1 }{\Delta m^2_{21} [(\cos2\theta_{12}-\Tilde{A})^2+\sin^2 2\theta_{12}]^{3/2}} \,.
\end{split}  
\end{equation}
which can be obtained from the previous ones with the substitution $m_1\leftrightarrow m_2$. The discussions about the effective mass splitting are unchanged in the case of $\eta_2$. On the other hand, we can observe that now in order to obtain a negative correction for $m_1\to0$ in the mixing angle, we need $\eta_2<0$. This allow for a successful data fit with a smaller $\theta_{12}$ for positive $\eta_2$ values.

Regarding $\eta_3$, we will only consider the case in which this parameter is real. The corrections to the solar parameters read
\begin{equation}
\begin{split}
    [\Delta m^2_{21,\rm{eff}}]_{\eta_3}&=2\bar{m}\, \eta_3\frac{\sin2\theta_{12}(m_1+m_2)}{\sqrt{(\cos2\theta_{12}-\Tilde{A})^2+\sin^2 2\theta_{12}}} \\
    [\cos2\theta_{12}^s]_{\eta_3}&=-2\bar{m}\, \eta_3 \, \sin^2 2\theta_{12} \frac{(\Tilde{A}-\cos2\theta_{12})(m_1+m_2) }{\Delta m^2_{21} [(\cos2\theta_{12}-\Tilde{A})^2+\sin 2\theta_{12}]^{3/2}} \,.
\end{split}  
\end{equation}
The correction to the mass splitting has again the same sign of the parameter $\eta_3$. This again prefers negative values for $\eta_3$ (which correspond to $\phi=\pi$ in the case of a complex parameter). On the other hand, the correction to the mixing angle has the opposite sign with respect to $\eta_3$ for neutrinos which undergo matter transition. In this case, as in the $\eta_1$ case, a fit with smaller $\theta_{12}$ values is viable if $\eta_3>0$.

\section{Solar data analysis in presence of diagonal scalar NSI parameters}

We now want to understand which values of the scalar NSI are allowed by solar neutrino measurements using the full-three flavor picture and for arbitrary $\eta$ values (that is, we are not following either of the simplifying approximations from the previous section).
Starting from the observed values for the electron neutrino disappearance probabilities by Borexino \cite{BOREXINO:2018ohr}
\begin{equation}
\begin{aligned}
P_{ee}(\mathrm{pp})&=0.57\pm 0.09 \\
P_{ee}(\mathrm{^7Be})&=0.53\pm0.05 \\
P_{ee}(\mathrm{pep})&=0.43\pm0.11 \\
P_{ee}(\mathrm{^8B})&=0.37\pm0.08
\end{aligned}
\label{eq:obsprob}
\end{equation}
and adding the spectral information on the $^8$B neutrinos by SNO \cite{SNO:2011hxd}\footnote{Additional solar neutrino information from the dedicated pp experiments SAGE, GALEX, and GNO \cite{GALLEX:1998kcz,SAGE:1999nng,Gavrin:2001sz,GNO:2000avz} and $^8$B measurements Super-Kamiokande \cite{Super-Kamiokande:2016yck} will add some additional information, but will not significantly enhance the constraint.} including uncertainties,
we then compute the $\chi^2$ function defined as
\begin{equation}
    \chi^2=\sum_{p}\frac{(P_{obs}^p-P(\theta_{12},\Delta m_{21}^2, \eta,\phi,m)^p)^2}{(\sigma^p_{obs})^2}\,,
\end{equation}
where $p$ are all the neutrino components already mentioned, $P_{obs}^p$ are the observed probabilities with their uncertainties $(\sigma^p_{obs})^2$ in eq.~\ref{eq:obsprob} and $P(\theta_{12},\Delta m_{21}^2,\eta,\phi,m)$ are the probabilities computed in presence of scalar NSI using the exact probabilities from eq.~\ref{3flav} fixing $\theta_{13}$ and the other oscillation parameters in the atmospheric sector ($\theta_{23}$, $\Delta m^2_{31}$) fixed to their best fits (see Tab.~\ref{Tabprob} \cite{Esteban:2020cvm}) while setting CP violation to 0. We have checked that in the solar oscillation regime the effect of the atmospheric parameters, of $\delta_{\rm CP}$, and of the reactor mixing angle is negligible even in presence of scalar NSI.
\begin{figure}
    \centering   
    \includegraphics[width=4.5cm,height=4.5cm]{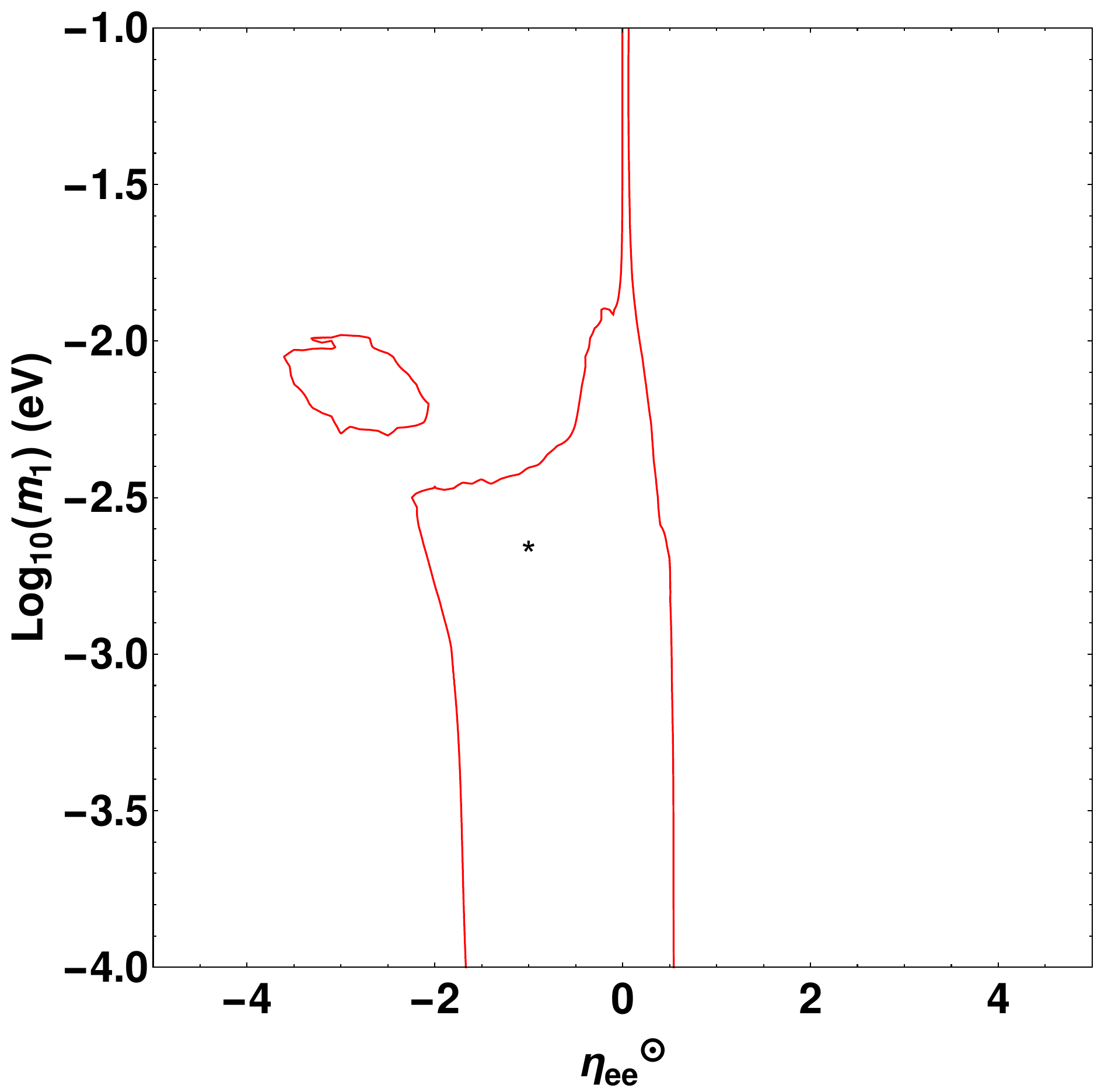}
    \includegraphics[width=4.5cm,height=4.5cm]{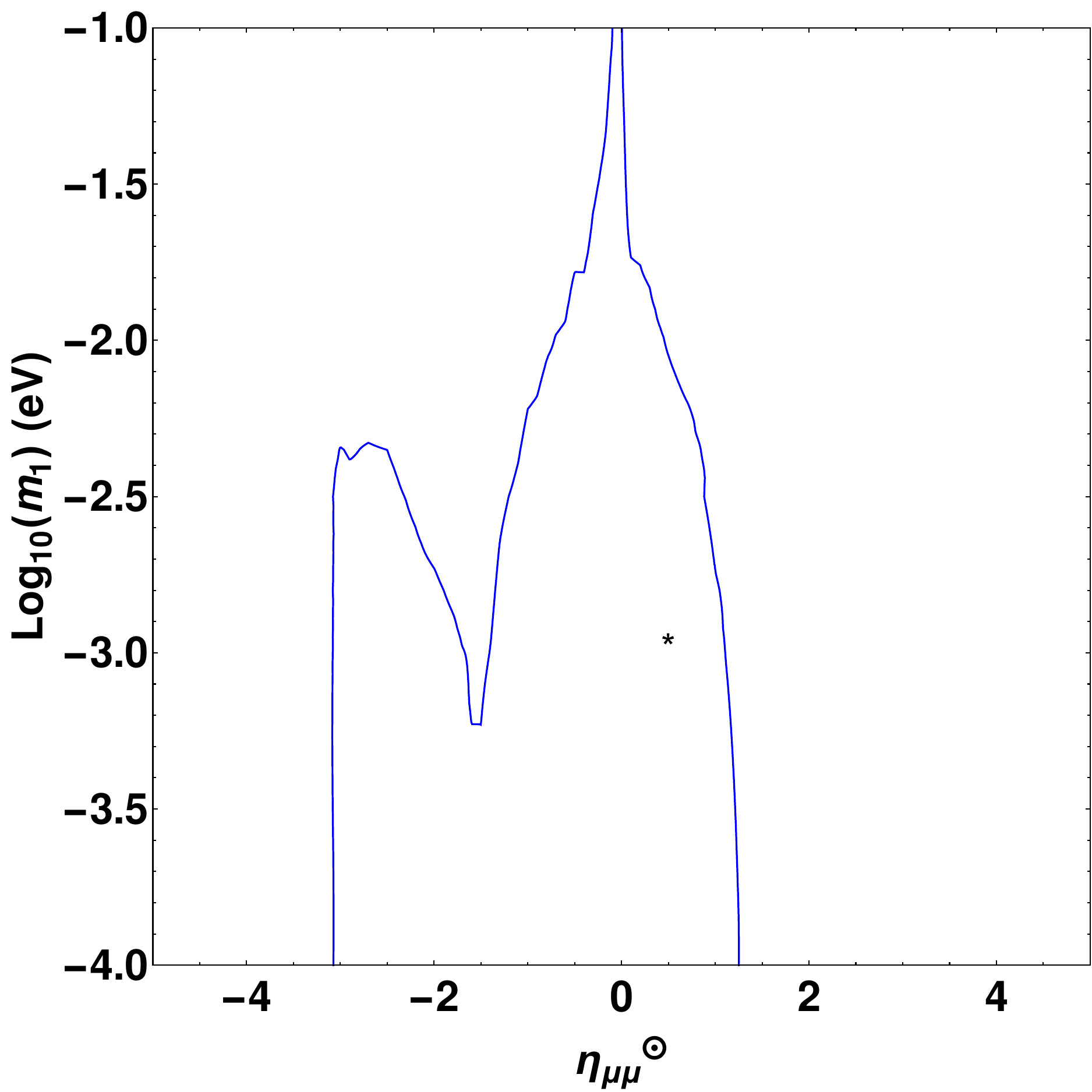}
    \includegraphics[width=4.5cm,height=4.5cm]{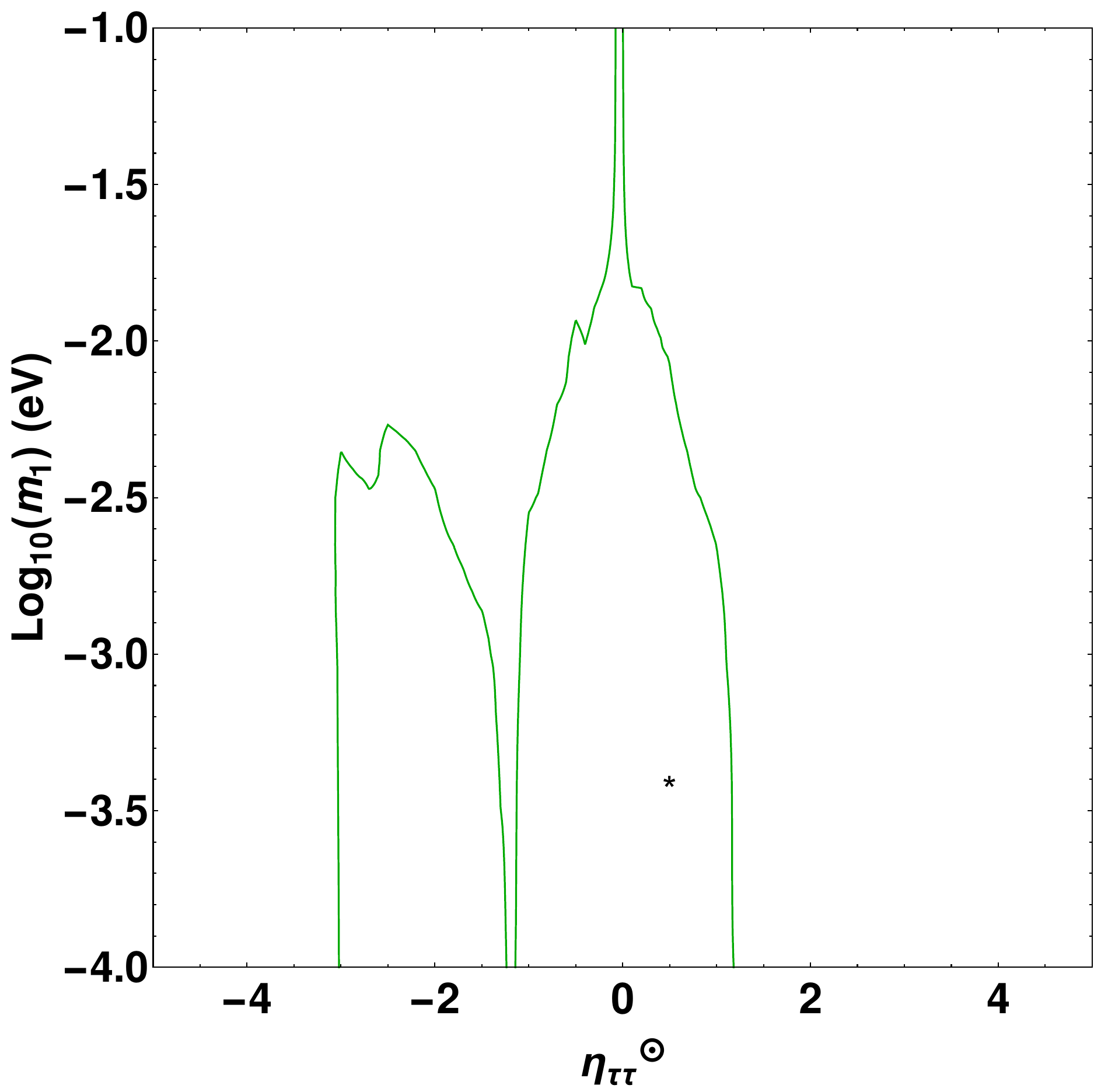}
    \caption{90\% CL contours in the $\eta_{\alpha\alpha}^\odot$ vs $m_1$ planes for the three diagonal scalar NSI parameters using solar neutrino data after minimizing over the relevant oscillation parameters. The stars depict the best fit points.}
    \label{fig:diagonalchi}
\end{figure}

In Fig.~\ref{fig:diagonalchi} we show the 90\% CL allowed regions at 2 degrees of freedom in the $\eta^\odot_{\alpha\alpha}-m_1$ planes marginalizing\footnote{For $\Delta m_{21}^2$ we consider only values in the range $[10^{-5}-10^{-4}]$ $eV^2$.} over the standard solar oscillation parameters: $\theta_{12}$ and $\Delta m^2_{21}$ while keeping the other sNSI parameters fixed to zero. It is clear that if the lightest neutrino mass is large ($m_1>10^{-2}$ eV), the scalar NSI parameters are strongly constrained. Indeed, for those values the new physics effects would be too large even for small $\eta_{\alpha\alpha}^\odot$ to reproduce the experimental measurements. However, if the lightest neutrino mass is small enough, data could be still reproduced if $\eta_{ee}^\odot\in[-2,0.5]$, $\eta_{\mu\mu,\tau\tau}^\odot\in[-3,1]$. In the valid approximation $\eta_1\to\eta_{ee}^\odot$ and $\eta_2\to\eta_{ee}^\odot$ it is possible to see that, as expected from the results shown in Sec.~\ref{sec:analytics}, negative values of the $\eta_{\alpha\alpha}^\odot$ are preferred since, in the case of large positive values, the matter transition would not occur for $^8$B neutrinos.

In agreement with \cite{Ge:2018uhz}, we find the negative best fit value $\eta_{ee}^\odot=-0.57$, for a lightest neutrino mass of roughly $m_1\sim4\times10^{-3}$ eV. On the other hand, the preferred values for $\eta_{\mu\mu}^\odot$ and $\eta_{\tau\tau}^\odot$ are positive. We now want to understand the effect of the presence of the scalar NSI parameters on the $\theta_{12}$ and $\Delta m_{21}^2$ measurements performed using solar neutrino data.
\begin{figure}
    \centering   
    \includegraphics[width=4.5cm,height=4.5cm]{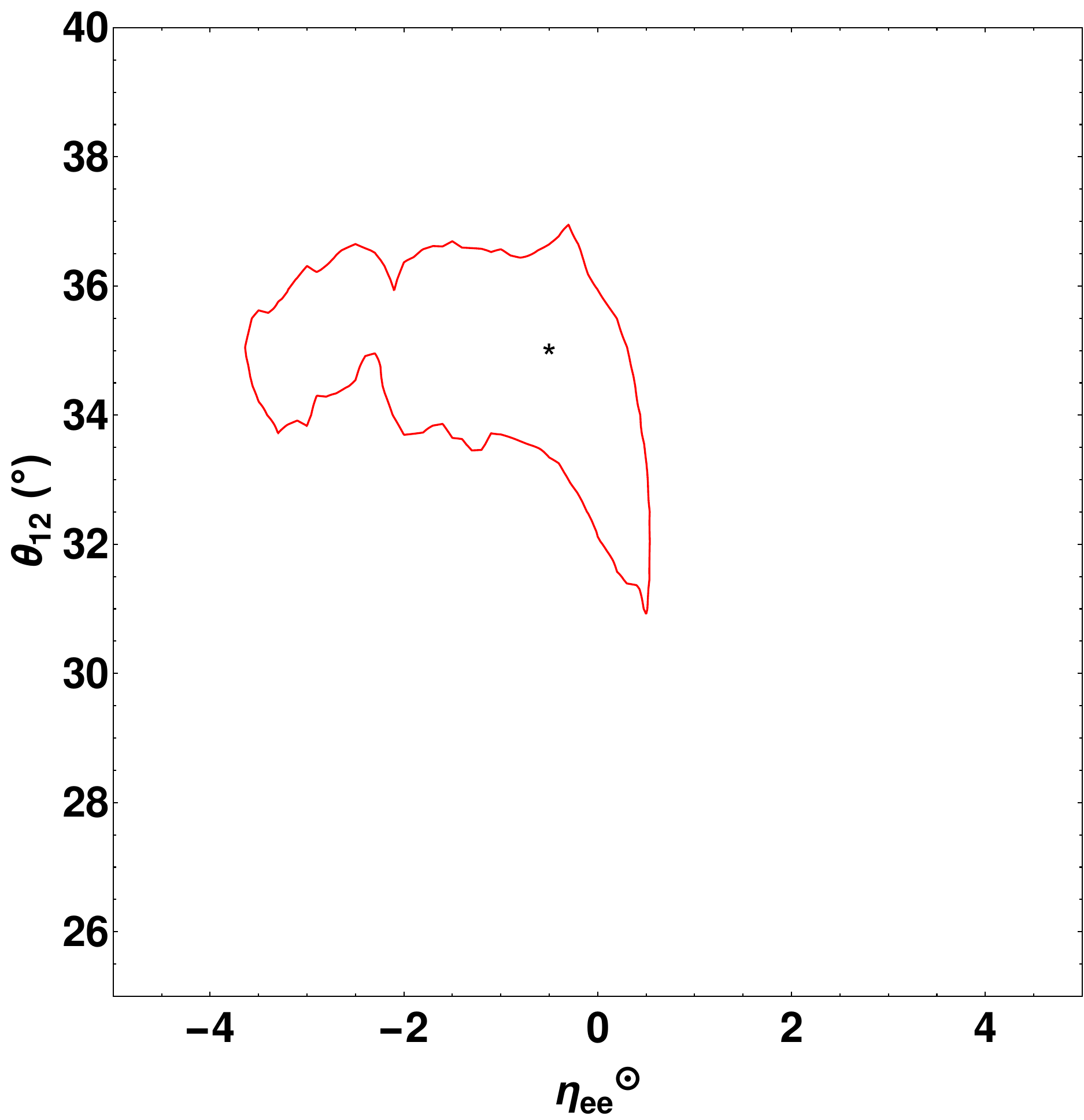}
    \includegraphics[width=4.5cm,height=4.5cm]{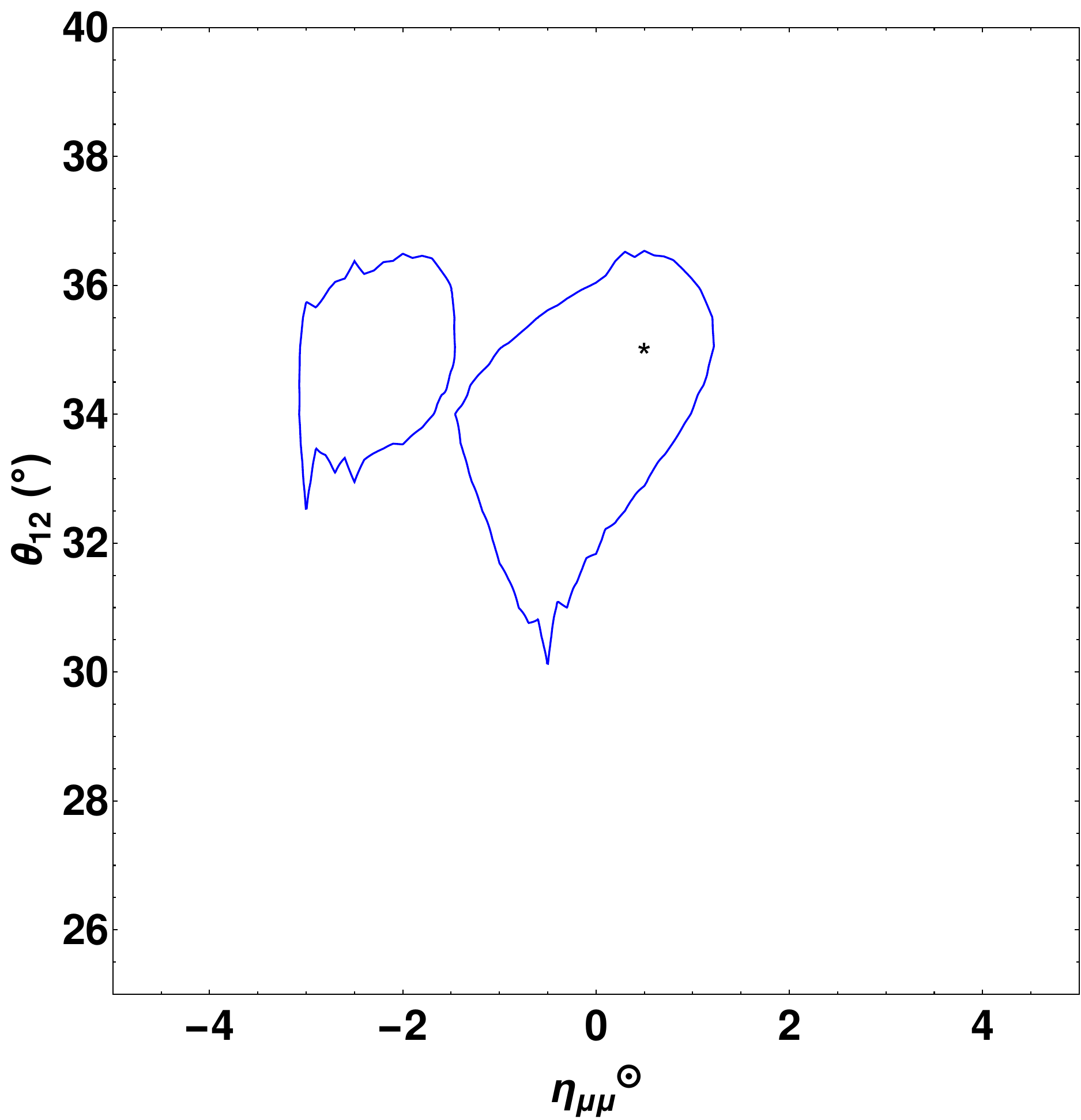}
    \includegraphics[width=4.5cm,height=4.5cm]{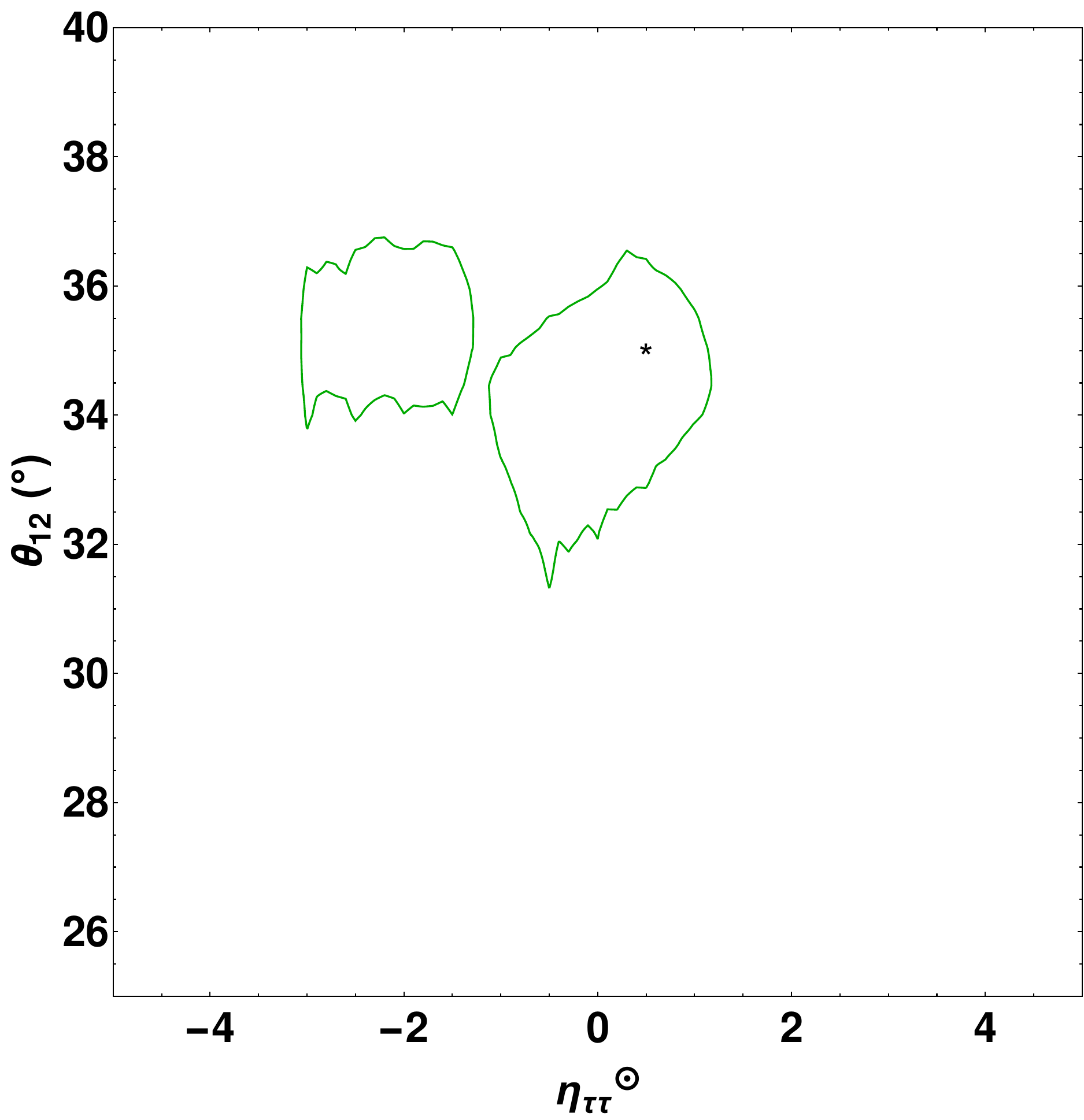}
    \caption{Same as Fig.~\ref{fig:diagonalchi} but in the $\eta_{\alpha\alpha}^\odot$-$\theta_{12}$ planes.}
    \label{fig:diagonalchi_th12}
\end{figure}
In Fig.~\ref{fig:diagonalchi_th12} we show the 90\% CL allowed regions in the $\eta^\odot_{\alpha\alpha}-\theta_{12}$ planes marginalizing over the solar mass splitting and the lightest neutrino mass. It is clear that the presence of non-zero $\eta_{\alpha\alpha}^\odot$ does not deteriorate the limits that the solar experiments may put on the solar mixing angle. Indeed, the widest allowed range in $\theta_{12}$ can be observed for all the three diagonal parameters when $\eta_{\alpha\alpha}^\odot\sim0$. This means that scalar NSI parameters do not affect $\theta_{12}$ measurement; the reason for that is that $pp$ neutrinos, which undergo "vacuum-like" oscillations, should not suffer from the effect of propagation scalar NSI. Thus, the $\theta_{12}$ measurement extracted from their observation remains robust. However, as expected from our previous discussion in Sec.~\ref{sec:analytics}, smaller values of the mixing angle can be allowed in the case of positive $\eta_{ee}^\odot$ or negative $\eta_{\mu\mu}^\odot$. We also checked that this happens for $m_1\to0$, in line with analytical expectations.

Finally, let us consider what happens to the determination of the mass splitting, keeping in mind that solar data can in principle bound this parameter only looking at the transition region between the vacuum and the matter dominated domains. 
\begin{figure}
    \centering   
    \includegraphics[width=4.5cm,height=4.5cm]{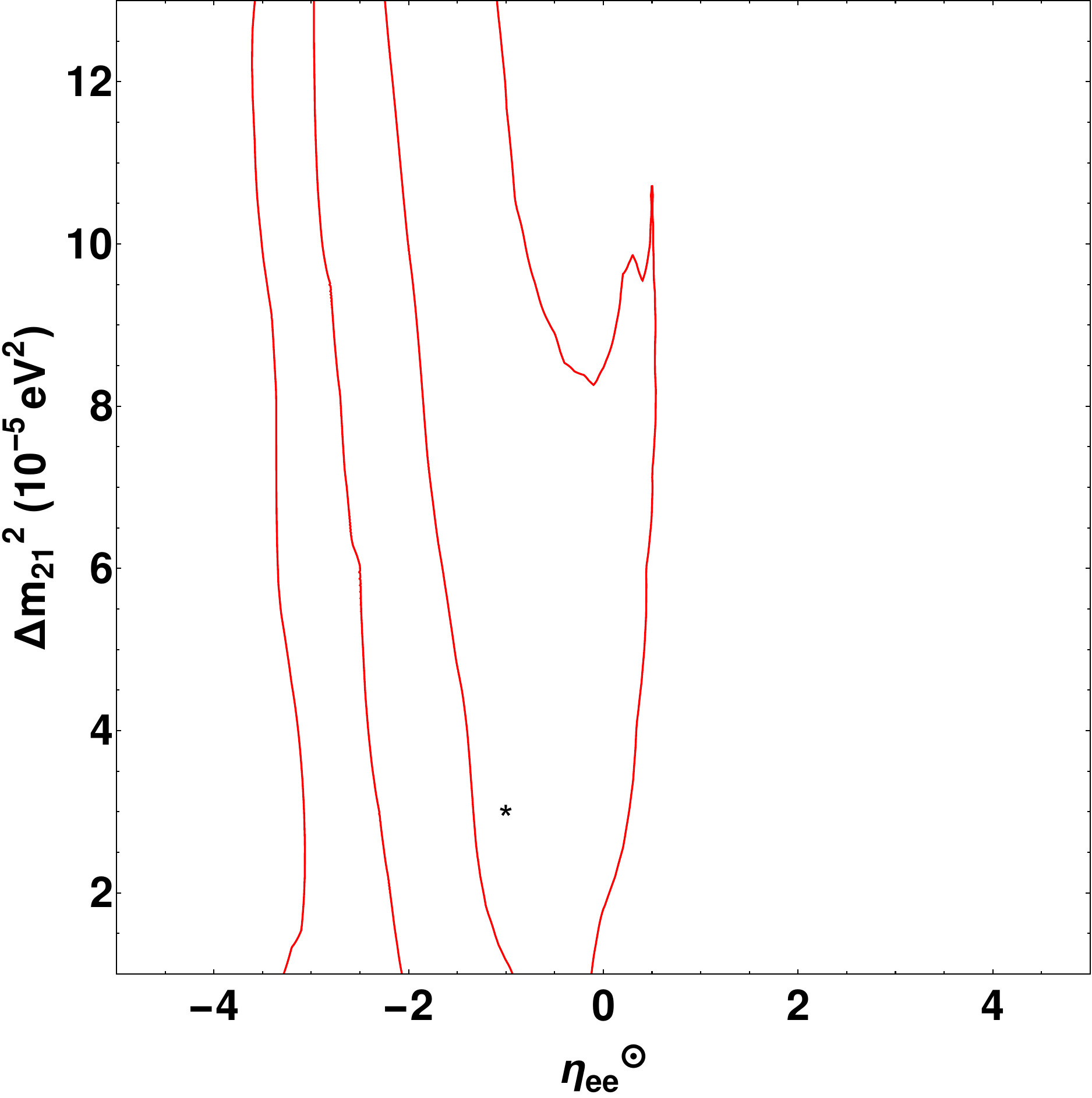}
    \includegraphics[width=4.5cm,height=4.5cm]{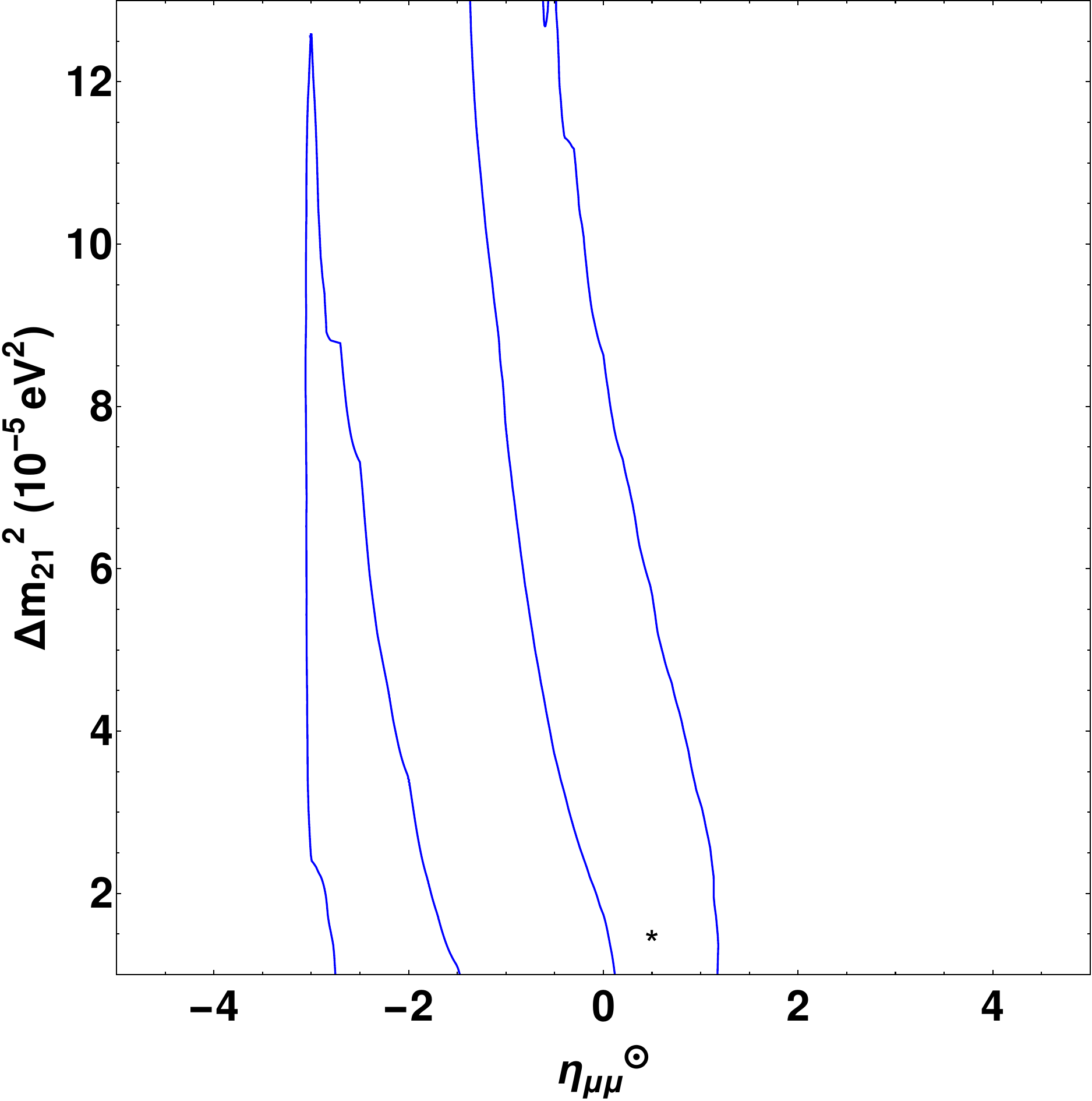}
    \includegraphics[width=4.5cm,height=4.5cm]{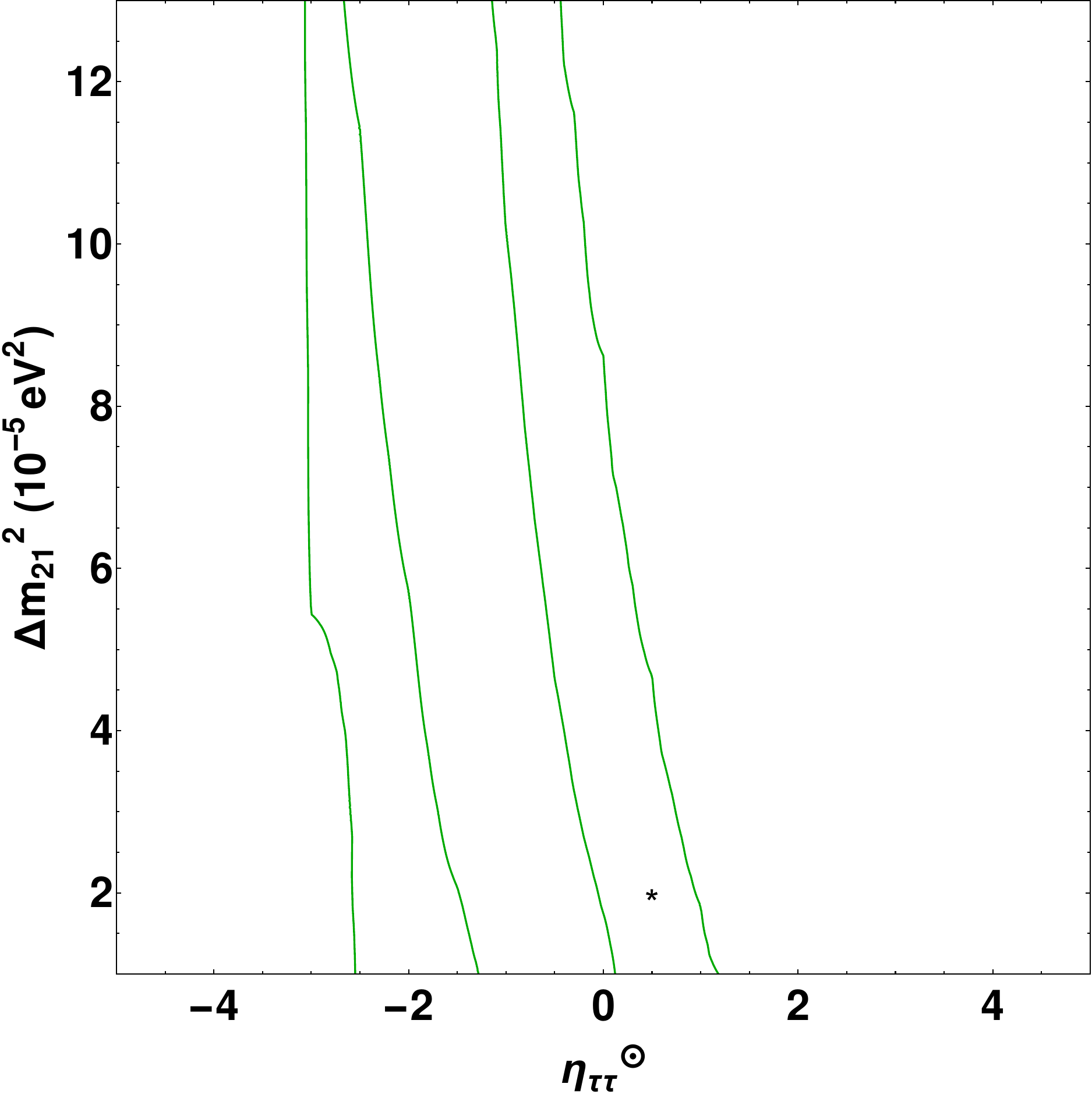}
    \caption{Same as Fig.~\ref{fig:diagonalchi} but in the $\eta_{\alpha\alpha}^\odot$-$\Delta m_{21}^2$ planes.}
    \label{fig:diagonalchi_dm}
\end{figure}
In Fig.~\ref{fig:diagonalchi_dm} we show the 90\% CL allowed regions in the $\eta^\odot_{\alpha\alpha}-\Delta m_{21}^2$ planes marginalizing over $\theta_{12}$ and the lightest neutrino mass. It is clear that the weak bounds on the solar mass splittings which could be determined using solar neutrino data are worsened by the scalar NSI. In fact, as already discussed, the structure of the neutrino Hamiltonian in presence of scalar NSI can fake the existence of the matter transition in the Sun even in presence of very low $\Delta m_{21}^2$. Indeed, as shown in Fig.~\ref{fig:diagonalchi_dm}, small values of the mass splittings are favorable, even though we checked that in the range $\Delta m_{21}^2\in[10^{-5}-10^{-4}] \, eV^2$, it is always possible to find a set of parameters for which $\Delta \chi^2$ is very low. Therefore, it is straightforward that the KamLAND-solar data tension in the context of the solar mass splitting could easily be explained by the presence of scalar NSI. 

\section{Solar data analysis in presence of non-diagonal scalar NSI parameters}

\begin{figure}
    \centering   \includegraphics[width=4.5cm,height=4.5cm]{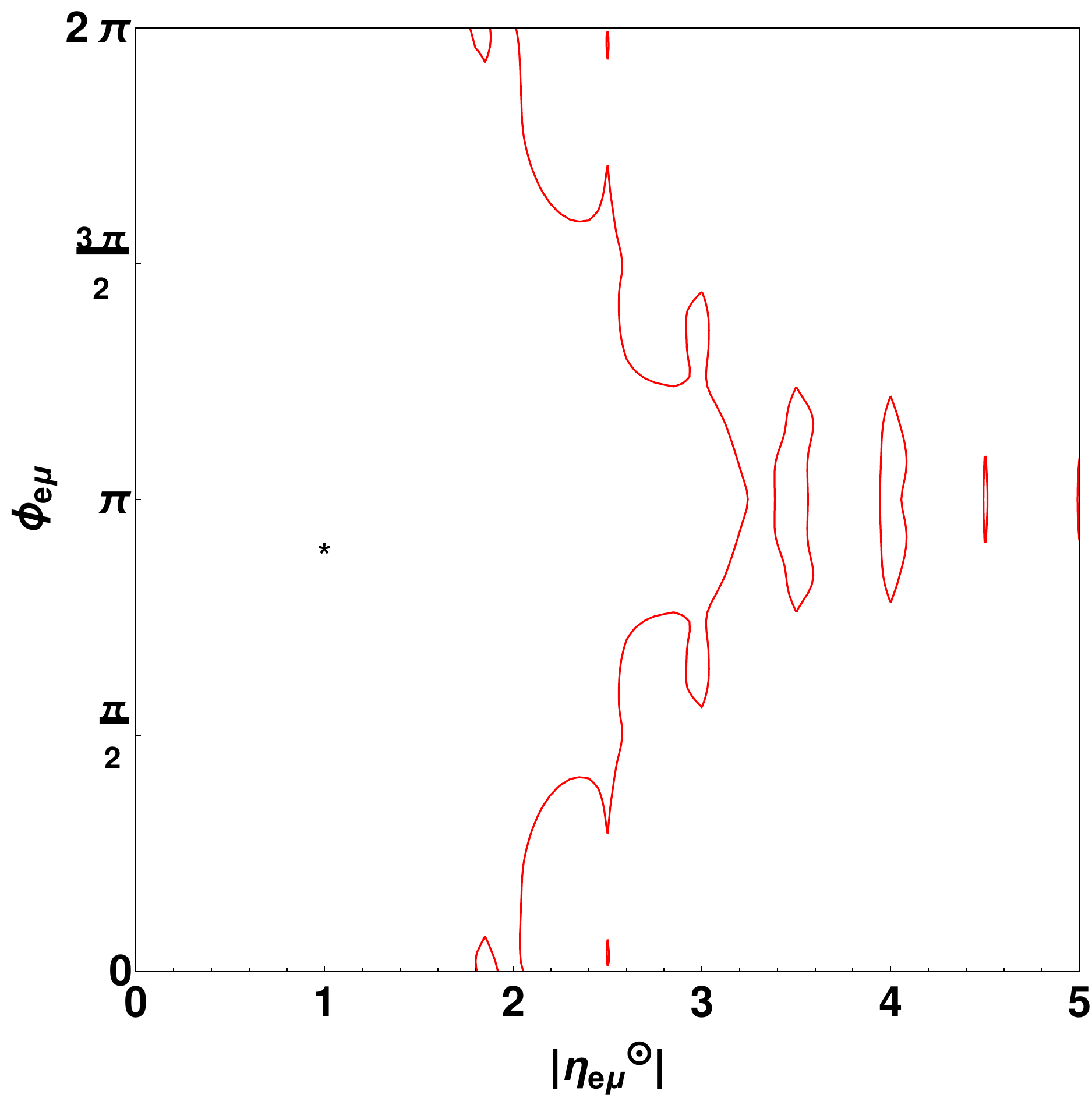}
    \includegraphics[width=4.5cm,height=4.5cm]{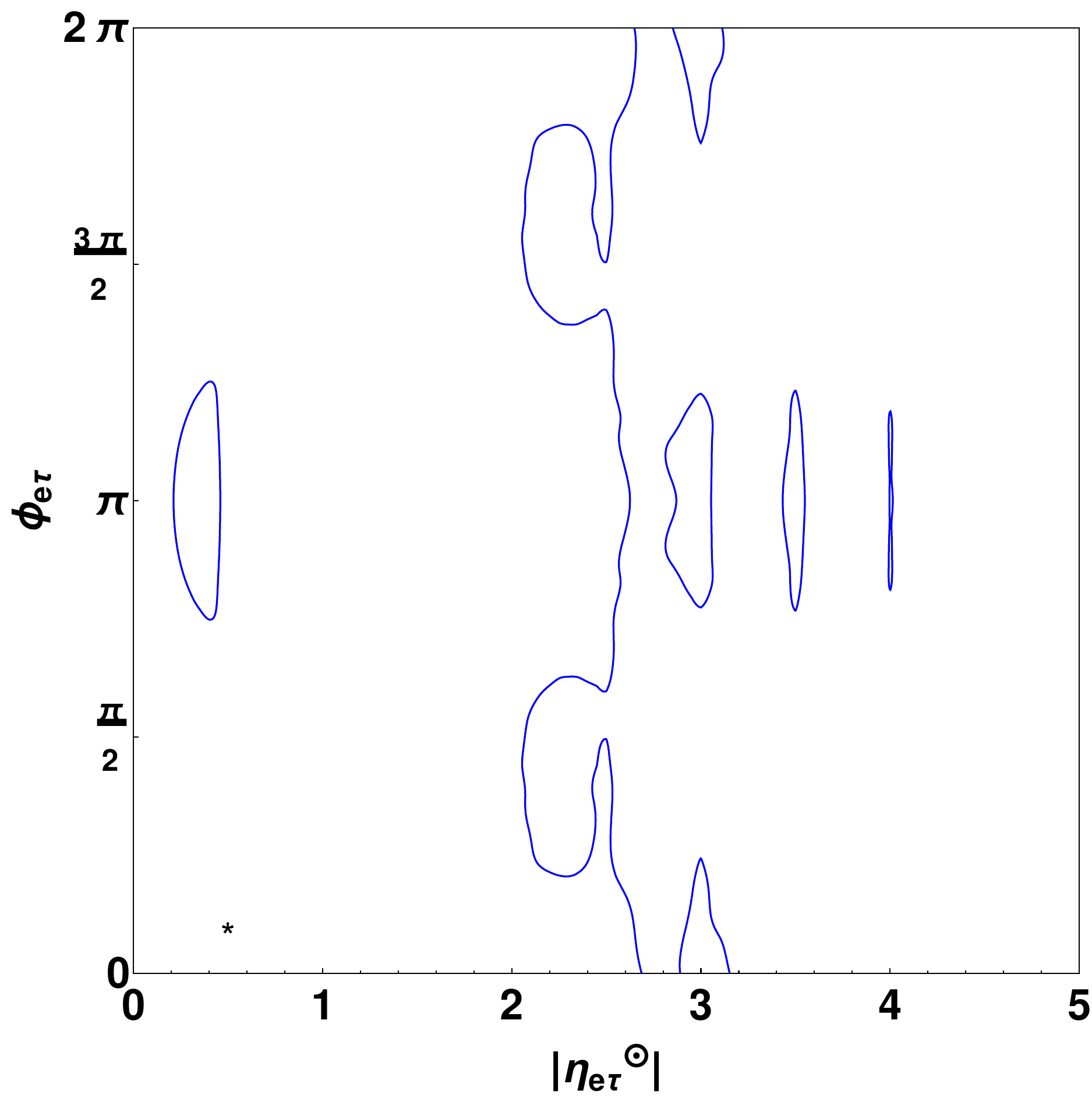}
    \includegraphics[width=4.5cm,height=4.5cm]{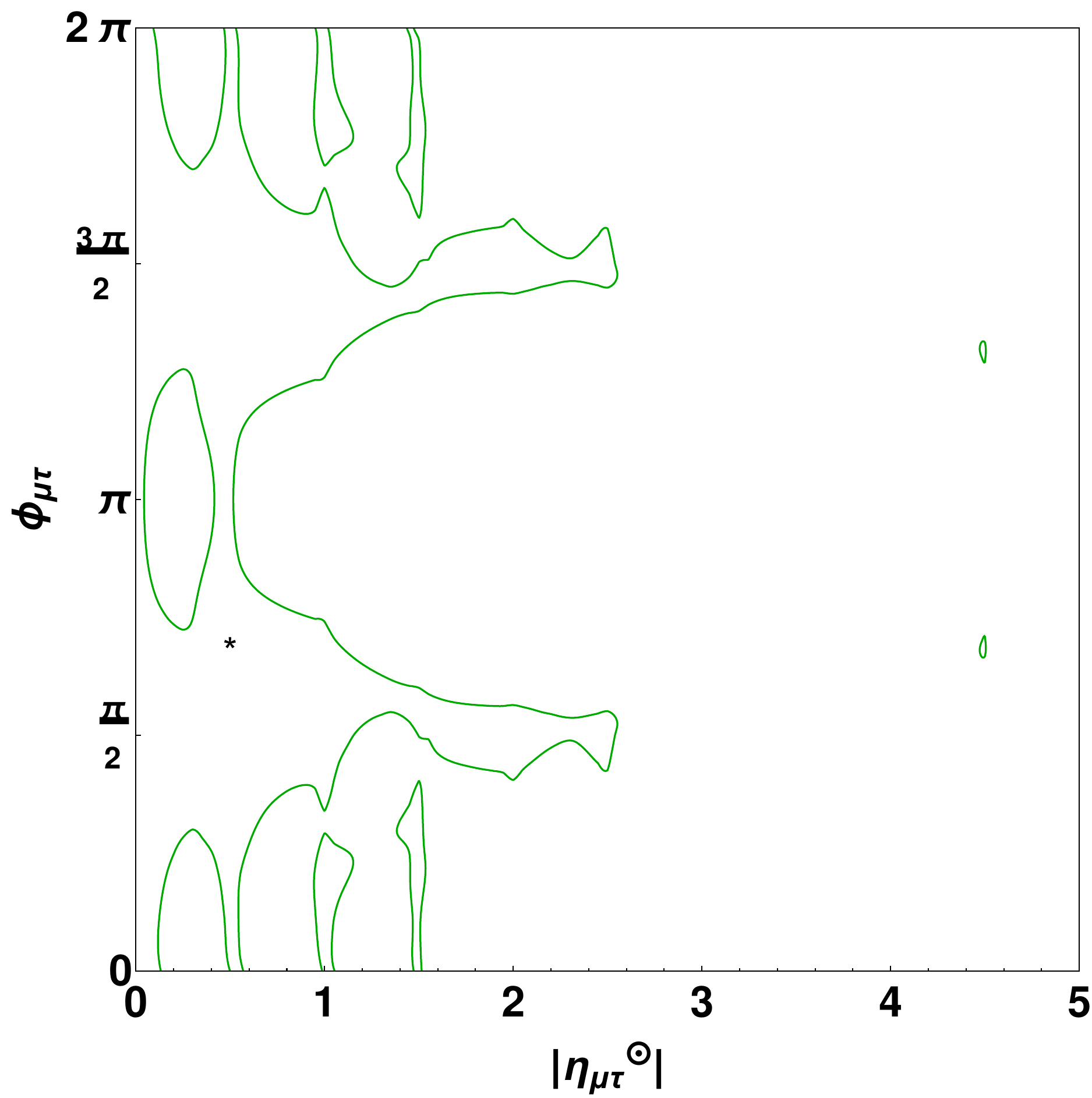}
    \caption{90\% CL contours in the $|\eta_{\alpha\beta}^\odot|$ vs $\phi_{\alpha\beta}$ planes for the three non-diagonal scalar NSI parameters.}
    \label{fig:nondiagonalchi}
\end{figure}

\begin{figure}
    \centering   \includegraphics[width=4.5cm,height=4.5cm]{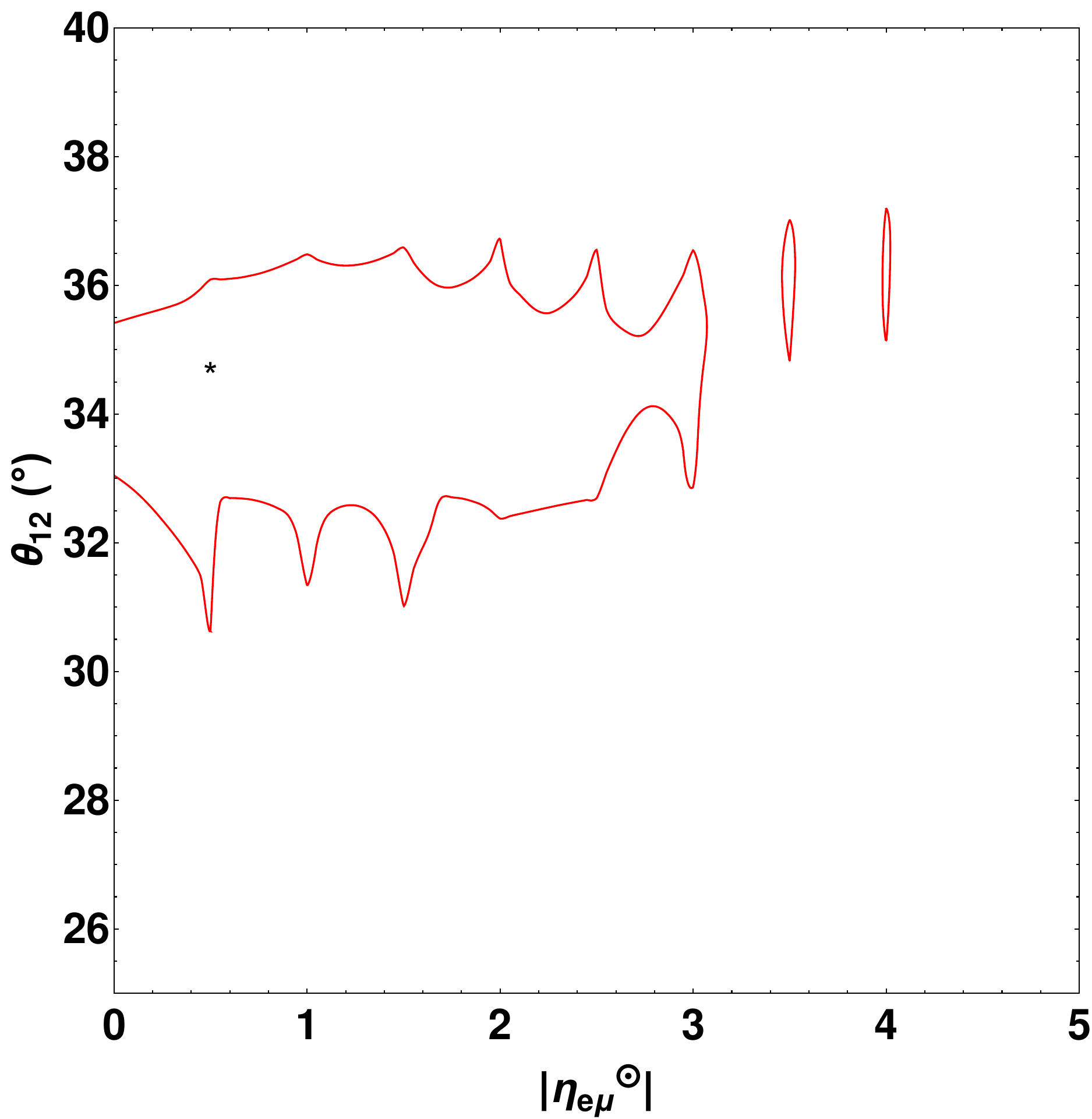}
    \includegraphics[width=4.5cm,height=4.5cm]{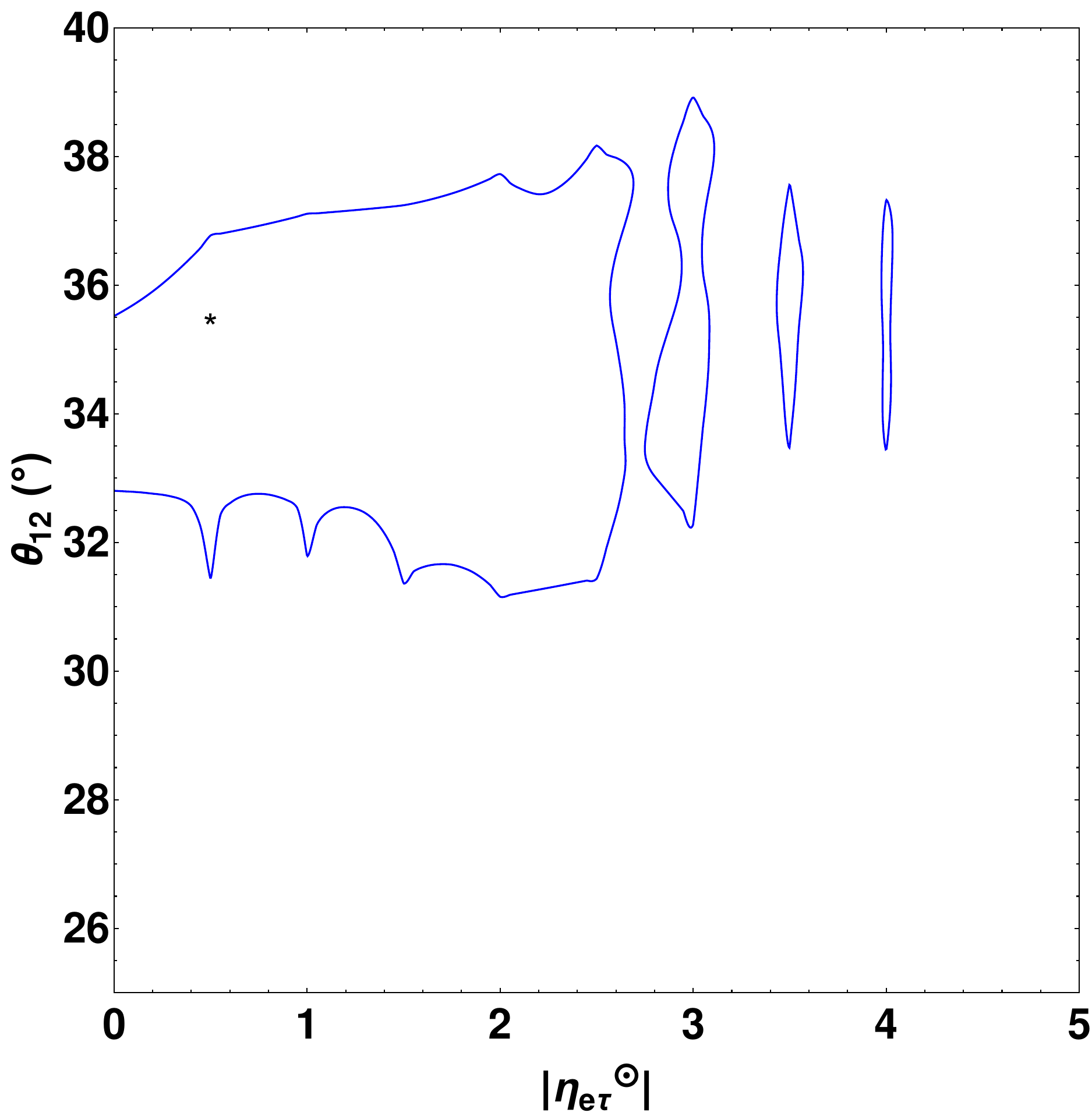}
    \includegraphics[width=4.5cm,height=4.5cm]{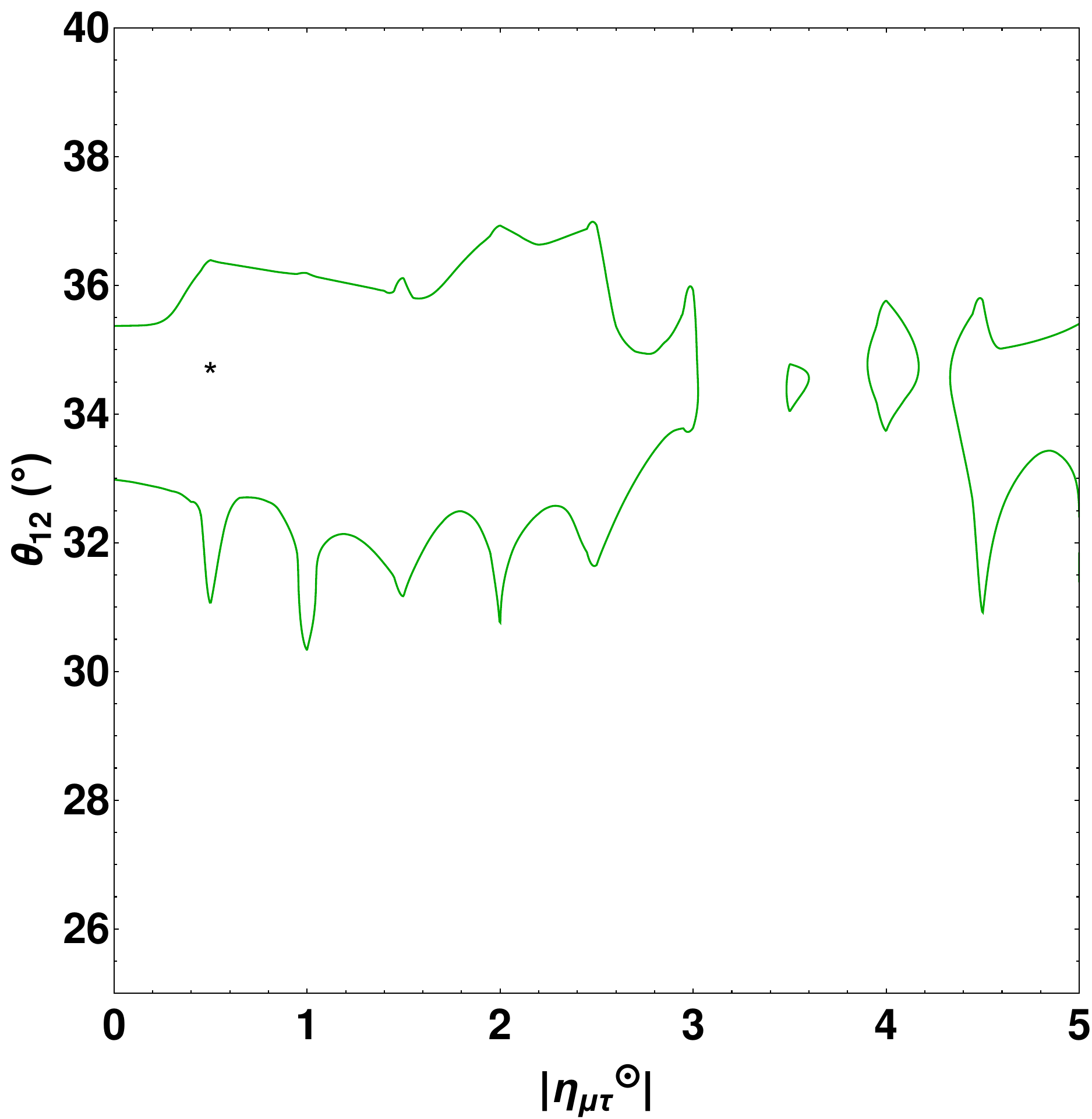}
    \caption{Same as Fig.~\ref{fig:nondiagonalchi} but in the $|\eta_{\alpha\beta}^\odot|$ vs $\theta_{12}$ planes.}
    \label{fig:nondiagonalchi_th12}
\end{figure}

\begin{figure}
    \centering   \includegraphics[width=4.5cm,height=4.5cm]{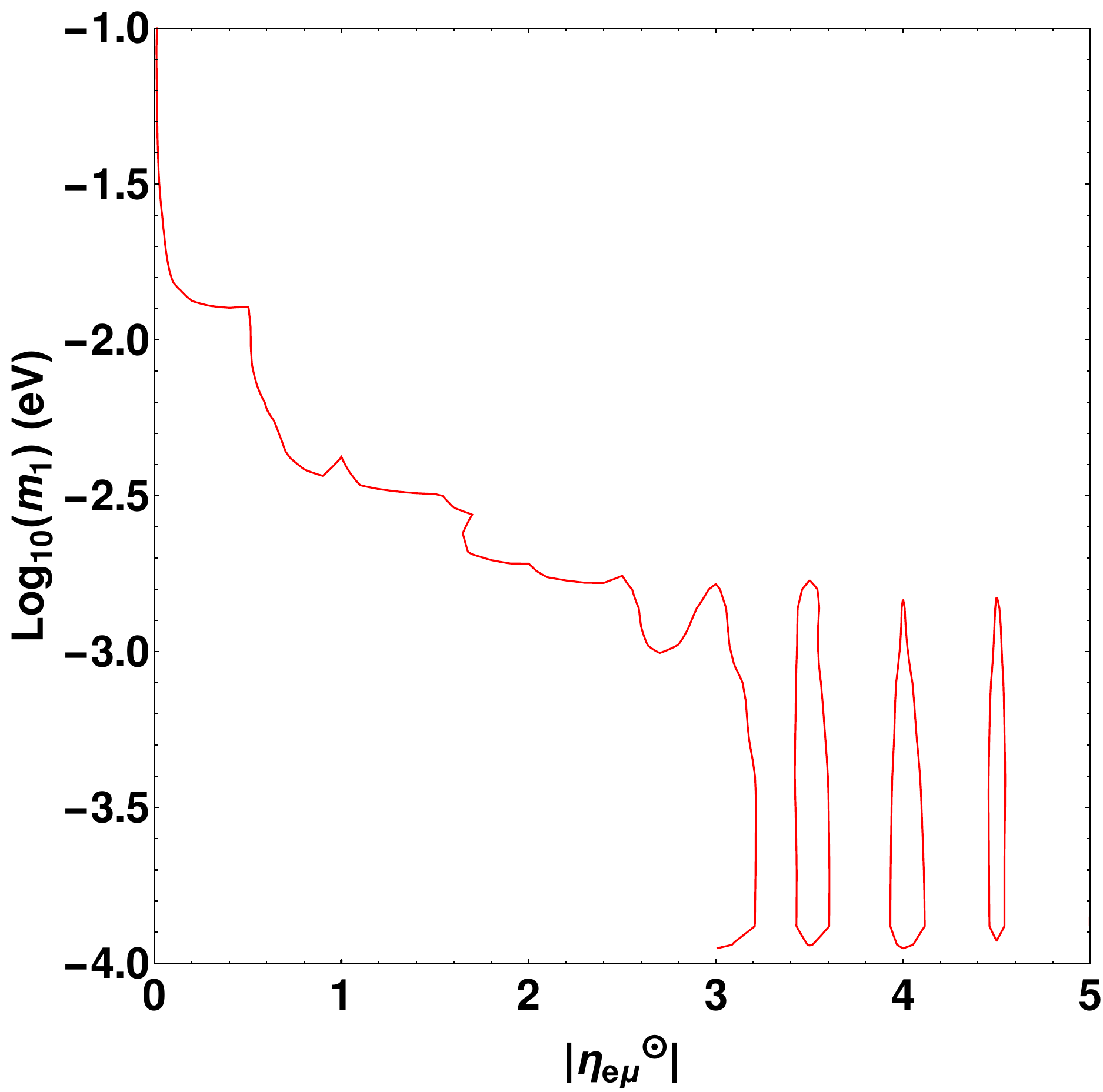}
    \includegraphics[width=4.5cm,height=4.5cm]{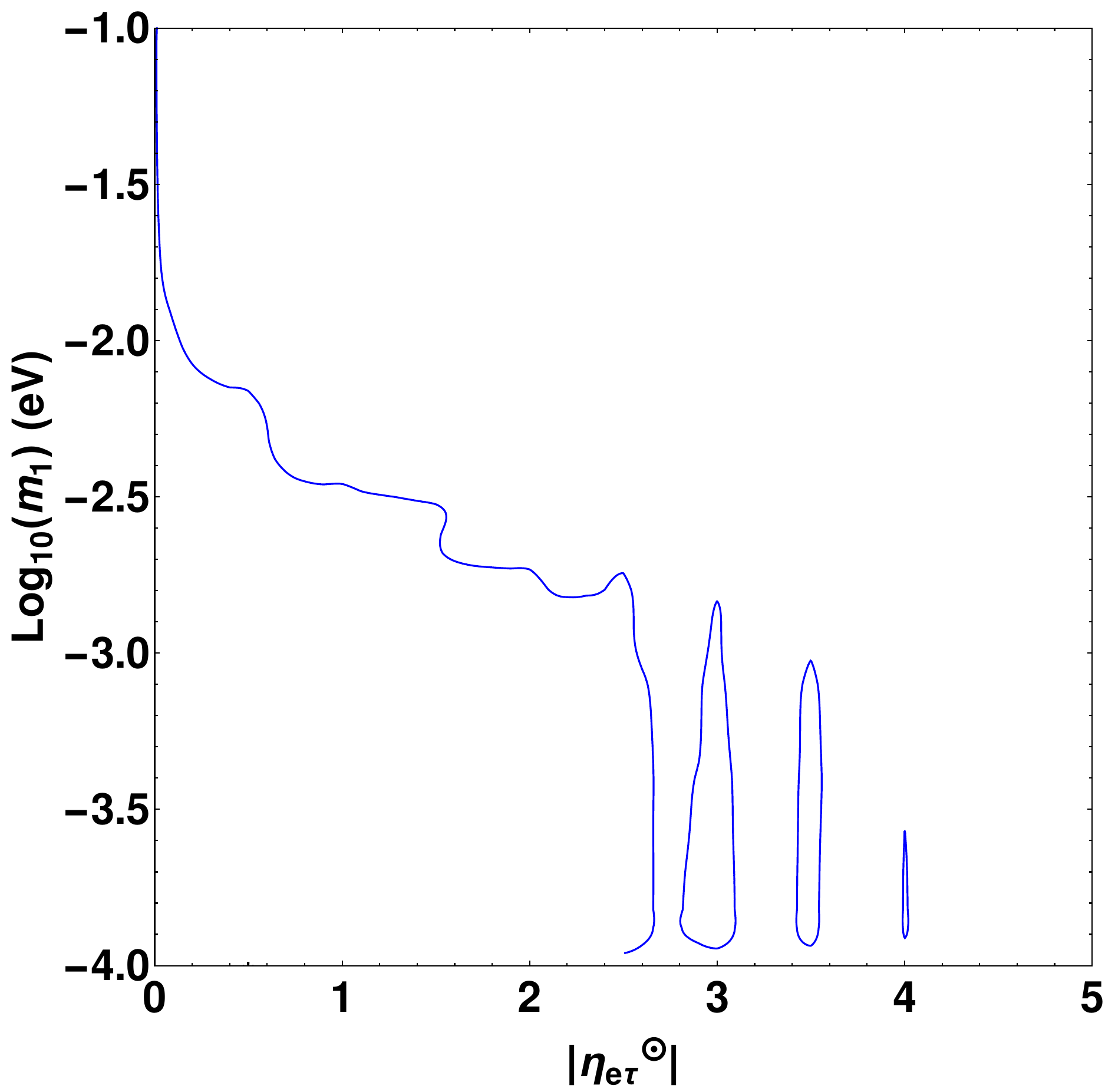}
    \includegraphics[width=4.5cm,height=4.5cm]{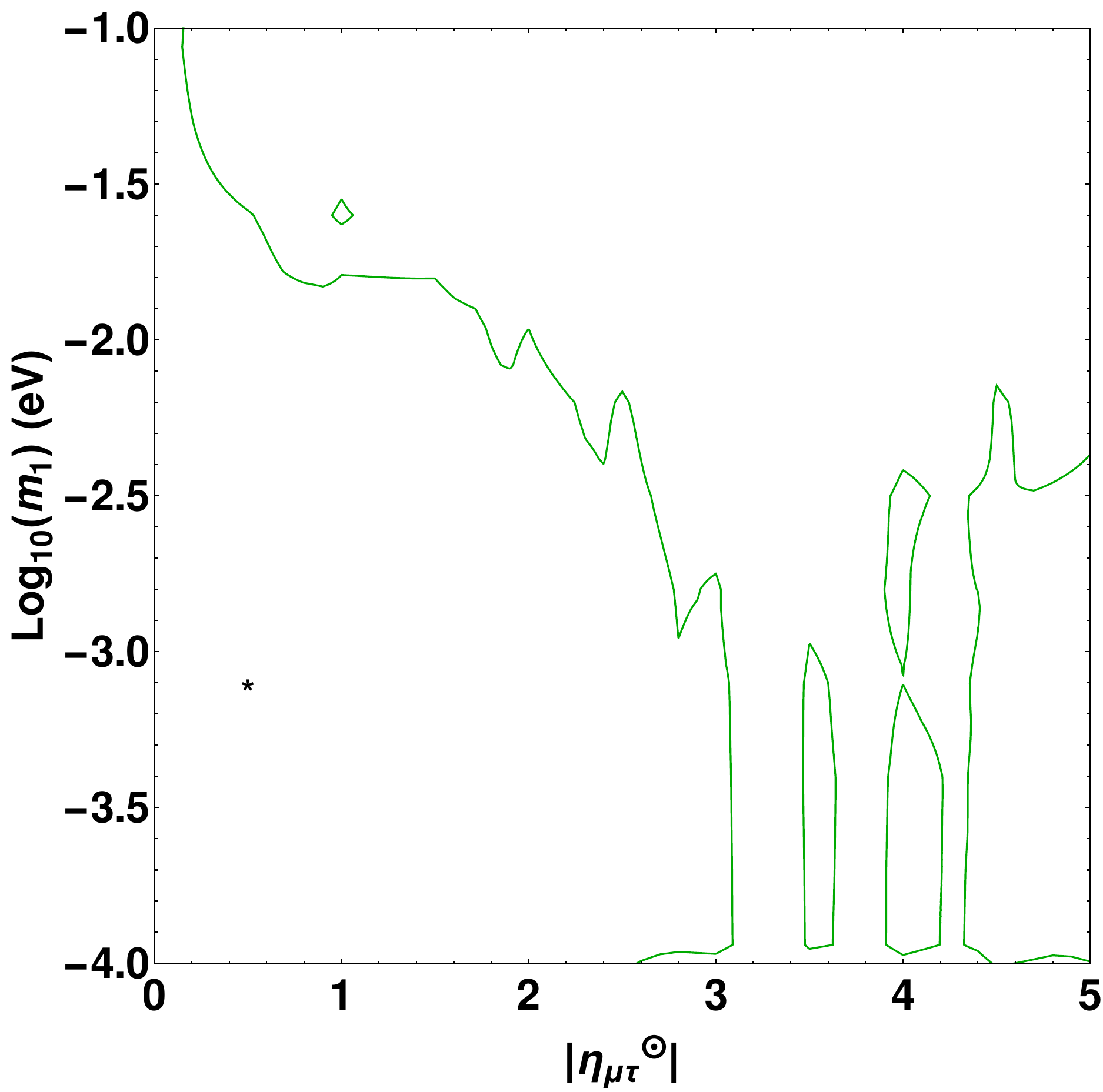}
    \caption{Same as Fig.~\ref{fig:nondiagonalchi} but in the $|\eta_{\alpha\beta}^\odot|$ vs $m_1$ planes.}
    \label{fig:nondiagonalchi_m1}
\end{figure}

We will explore now the effect of non-diagonal scalar NSI parameters on the solar neutrino oscillation. In this case we have another parameter which could in principle affect the probabilities, namely the phase of the non-diagonal parameters $\phi_{\alpha\beta}$. We show the 90\% allowed regions in the planes $|\eta_{\alpha\beta}^\odot|-\phi_{\alpha\beta}$ planes in Fig.~\ref{fig:nondiagonalchi}. 
In order to obtain this plot we marginalize over all the not-shown parameters, namely $\theta_{12}$, $\Delta m_{21}^2$ and the lightest neutrino mass $m_1$. In the left plane, dedicated to the first of the three non-diagonal parameter, $\eta_{e\mu}^\odot$, we can notice that the most favorable value of the phase is $\phi_{e\mu}\sim\pi$, for which also larger values of the modulus of the NSI parameter are allowed. Since the effect of $\eta_{e\mu}^\odot$ is similar to effect of $\eta_3$ discussed in Sec.~\ref{sec:analytics}, this results is in line with our previous considerations. For all the other values of the phase, we observe $|\eta_{e\mu}^\odot|\lesssim2$. A similar pattern can be seen when the effect of $\eta_{e\tau}^\odot$ is taken into account. On the other hand, $\eta_{\mu\tau}^\odot$ is more restricted and only small values are allowed, except when $\phi_{\mu\tau}\sim\pm\pi/2$. As in the case of the diagonal scalar NSI parameters, the solar mixing angle measurements performed using solar neutrino data is robust, as shown in Fig.~\ref{fig:nondiagonalchi_th12}.
Finally, we show the effect of increasing the lightest neutrino mass on the limits one would set on the non-diagonal scalar NSI parameters. From Fig.~\ref{fig:nondiagonalchi_m1} it is possible to observe that in the case of $\eta_{e\mu}^\odot$ and $\eta_{e\tau}^\odot$, the lowest $\Delta \chi^2$ is obtained when $m_1\to0$. However, large values of the modulus of these two parameters are still allowed if $m_1<5\times10^{-2}$ eV, similarly to what happens to diagonal scalar NSI parameters. On the other hand, $\eta_{\mu\tau}^\odot$ may slightly prefer larger values of the lightest neutrino mass and non zero values for $|\eta_{\mu\tau}^\odot|$ are allowed even in the case of large $m_1$. 

\section{The effect of the priors on the solar parameters}

We showed that the presence of scalar NSI negatively affects the determination of the solar mass splitting $\Delta m_{21}^2$ by solar neutrino data. The KamLAND experiment \cite{KamLAND:2013rgu}, however, provides a good measurement for this parameter.
Because sNSI at the levels discussed here will have no appreciable effect on reactor neutrinos even at long baselines, we can consider the KamLAND result as being completely independent of any effects of sNSI and as a measurement of only the regular oscillation parameters.
Moreover, the JUNO experiment \cite{Djurcic:2015vqa} is expected to measure with a sub-percent precision the solar parameters \cite{JUNO:2022mxj}. In order to study how the priors given by terrestrial experiments might change the allowed regions in the scalar NSI parameter spaces, we included the KamLAND result and expected JUNO sensitivity with 6 years of data as follows:
\begin{equation}
    \begin{split}
        \rm{KamLAND:} &\quad \Delta m_{21}^2=7.41\pm0.21\times10^{-5} \, \, \, \rm{eV}^2 \\
        \rm{JUNO:} & \quad \Delta m_{21}^2=7.53\pm0.024\times10^{-5} \, \, \, \rm{eV}^2, \, \, \, \sin^2\theta_{12}=0.307\pm0.0016
    \end{split}
\end{equation}

\begin{figure}
    \centering   \includegraphics[width=4.5cm,height=4.5cm]{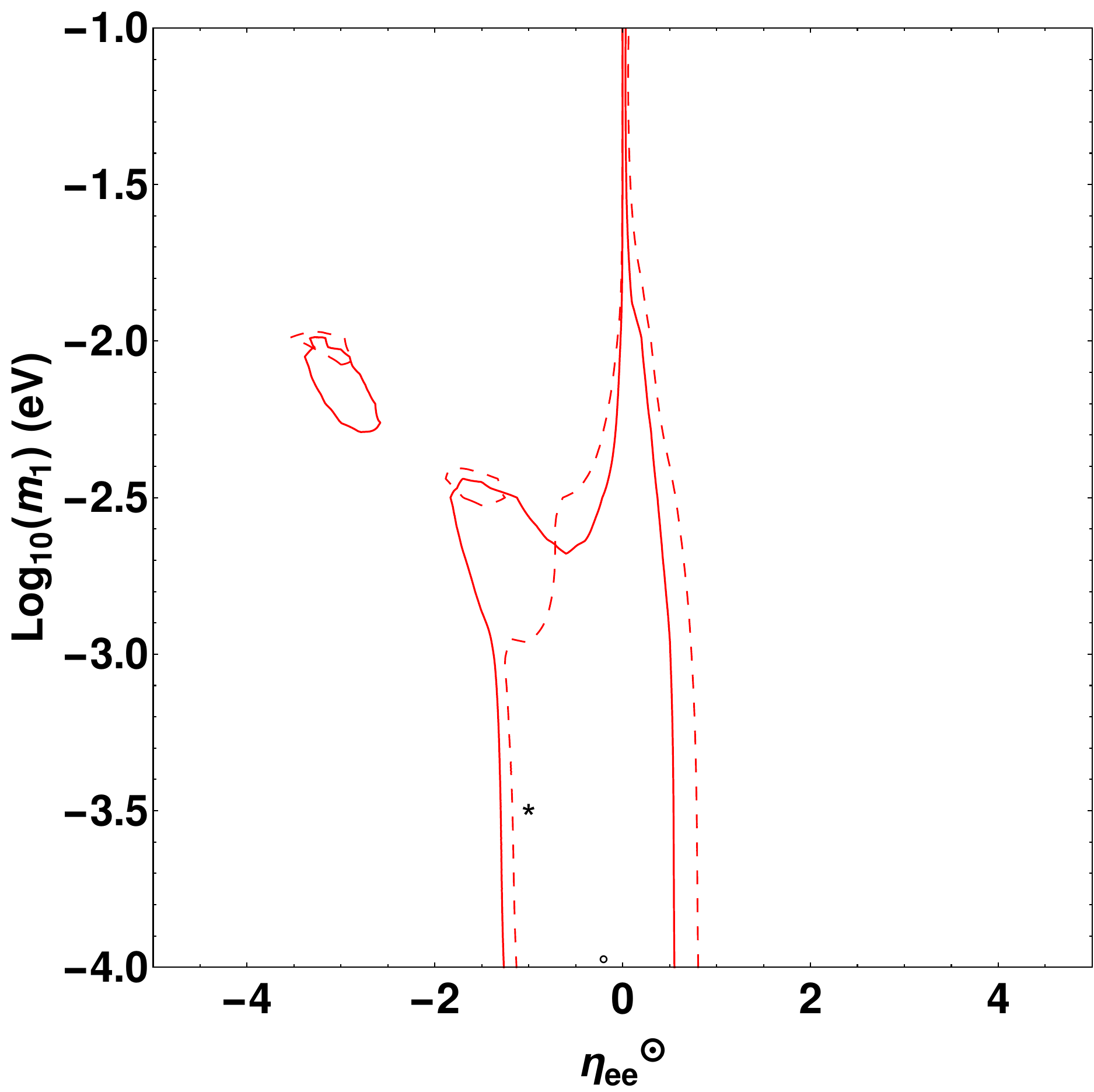}
    \includegraphics[width=4.5cm,height=4.5cm]{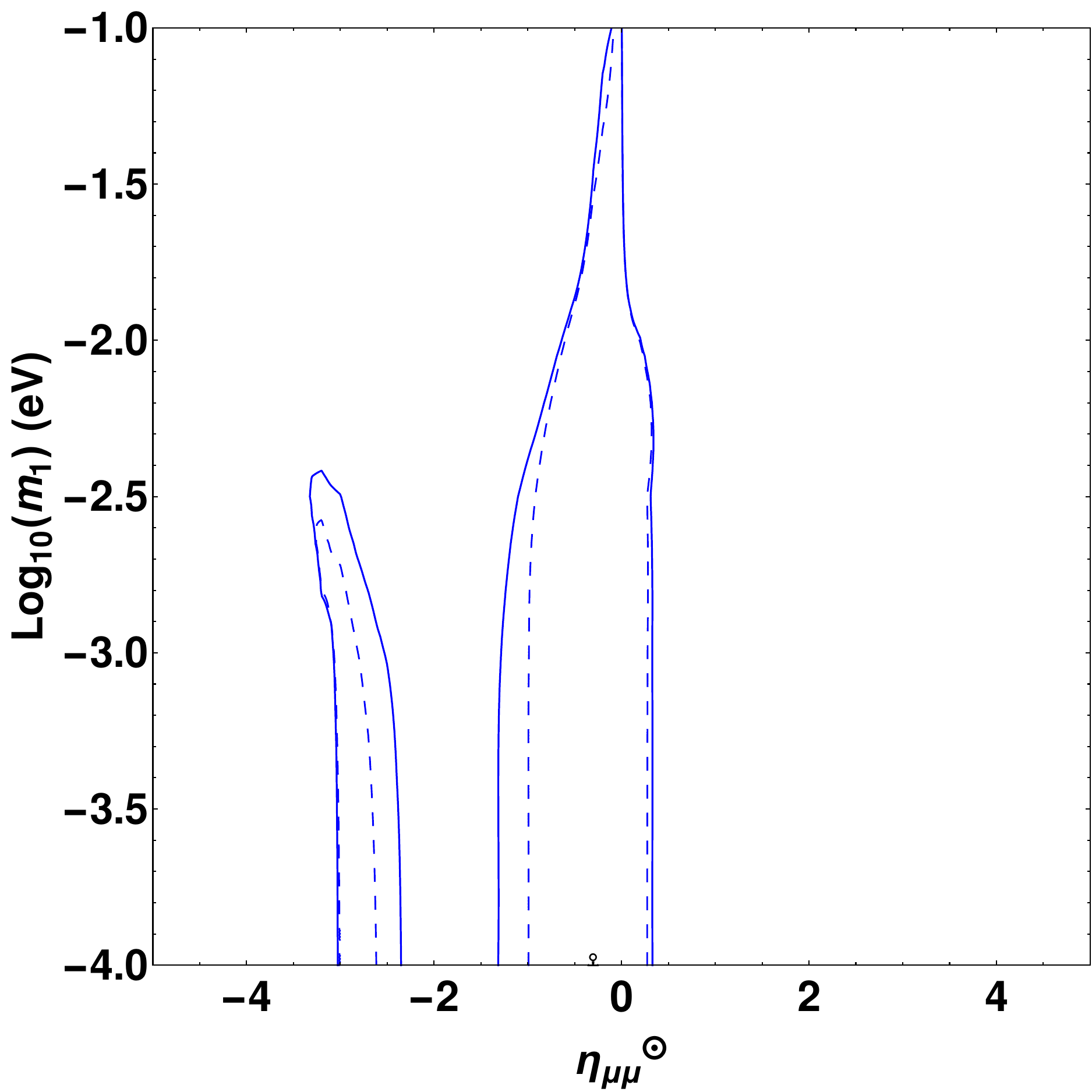}
    \includegraphics[width=4.5cm,height=4.5cm]{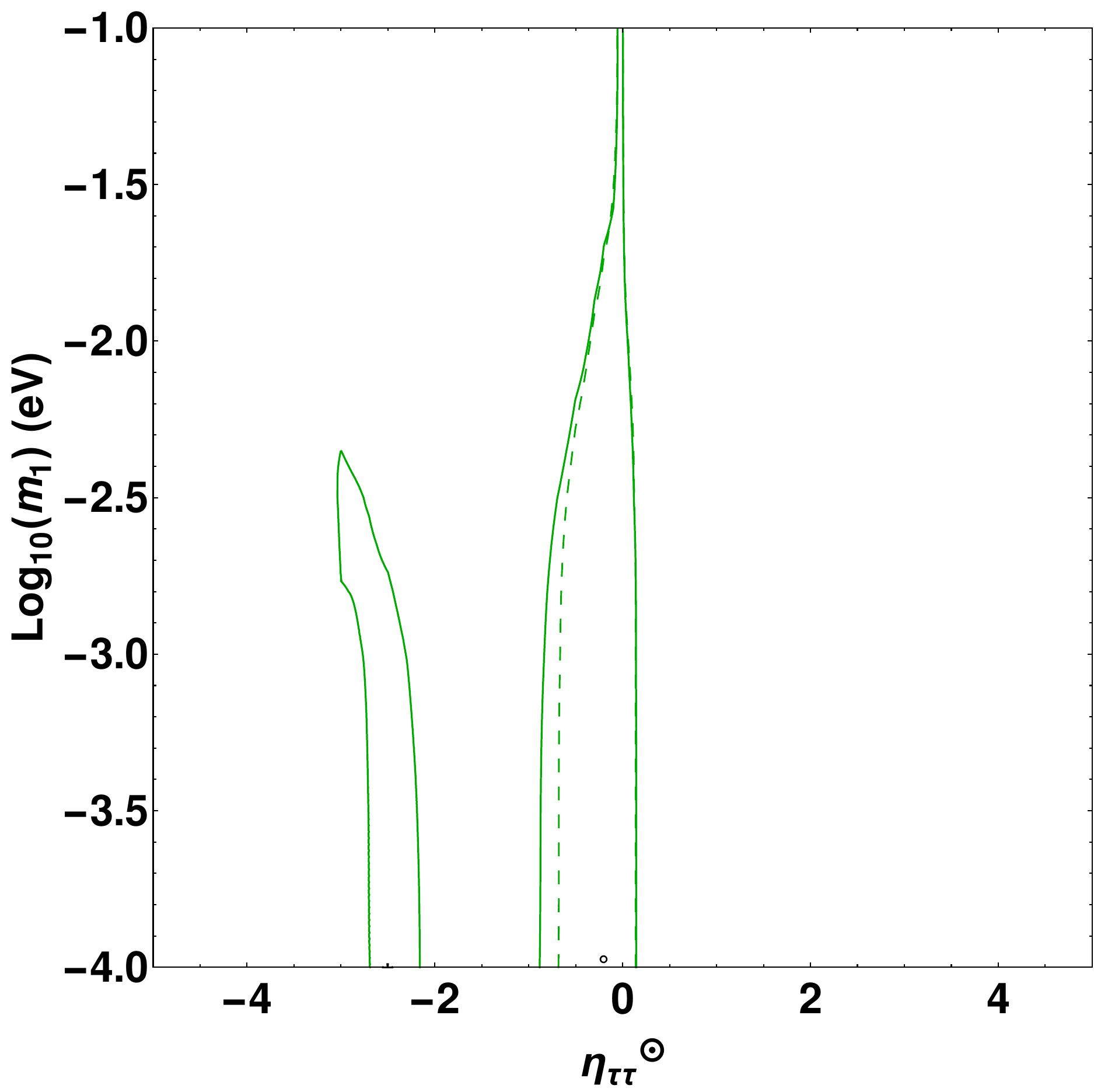}
    \caption{Same as Fig.~\ref{fig:diagonalchi} including the KamLAND (solid) and JUNO (dashed) prior on the solar parameters.}
    \label{fig:diagonalchi_m1_Kamland}
\end{figure}

\begin{figure}
    \centering   \includegraphics[width=4.5cm,height=4.5cm]{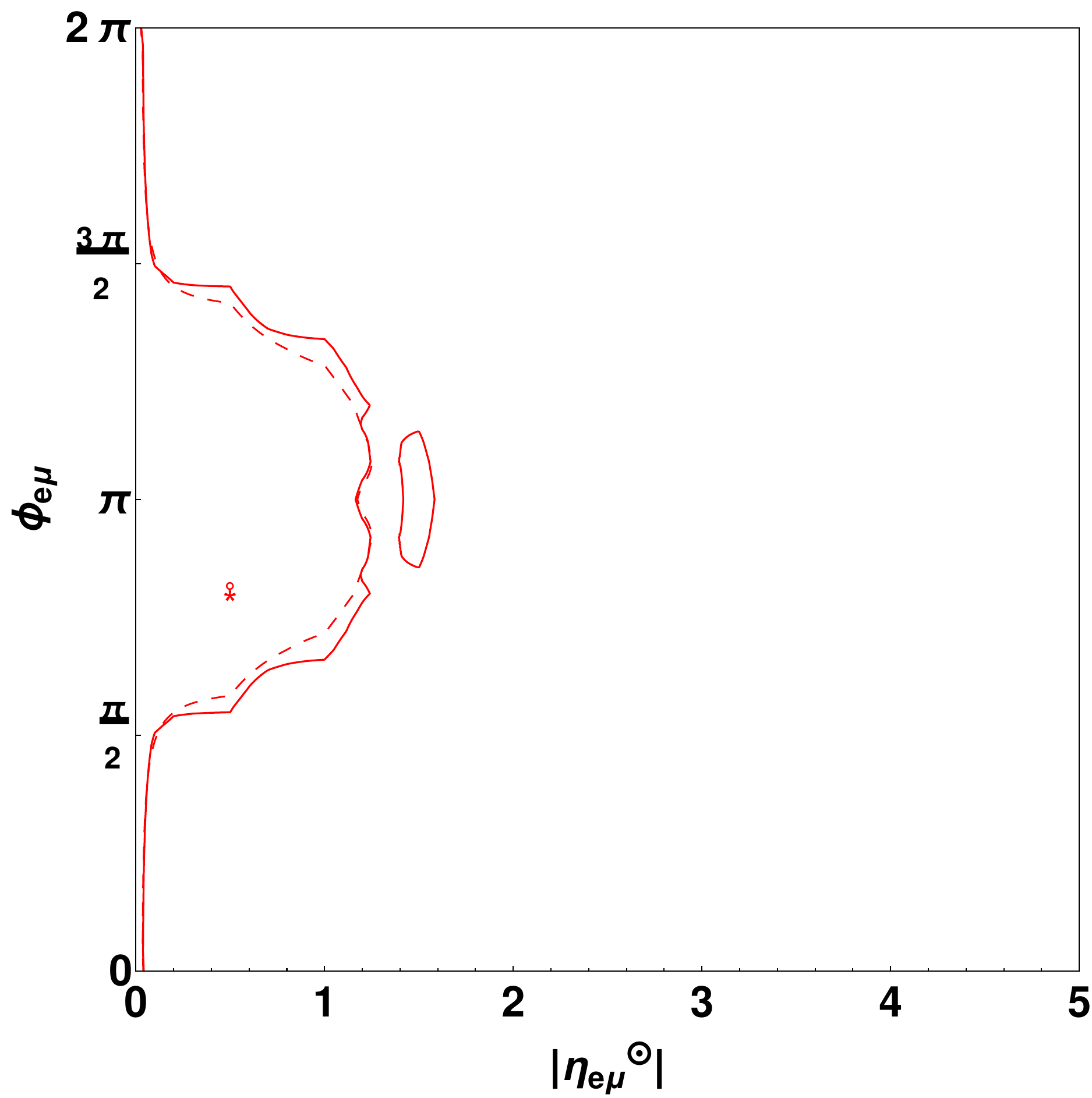}
    \includegraphics[width=4.5cm,height=4.5cm]{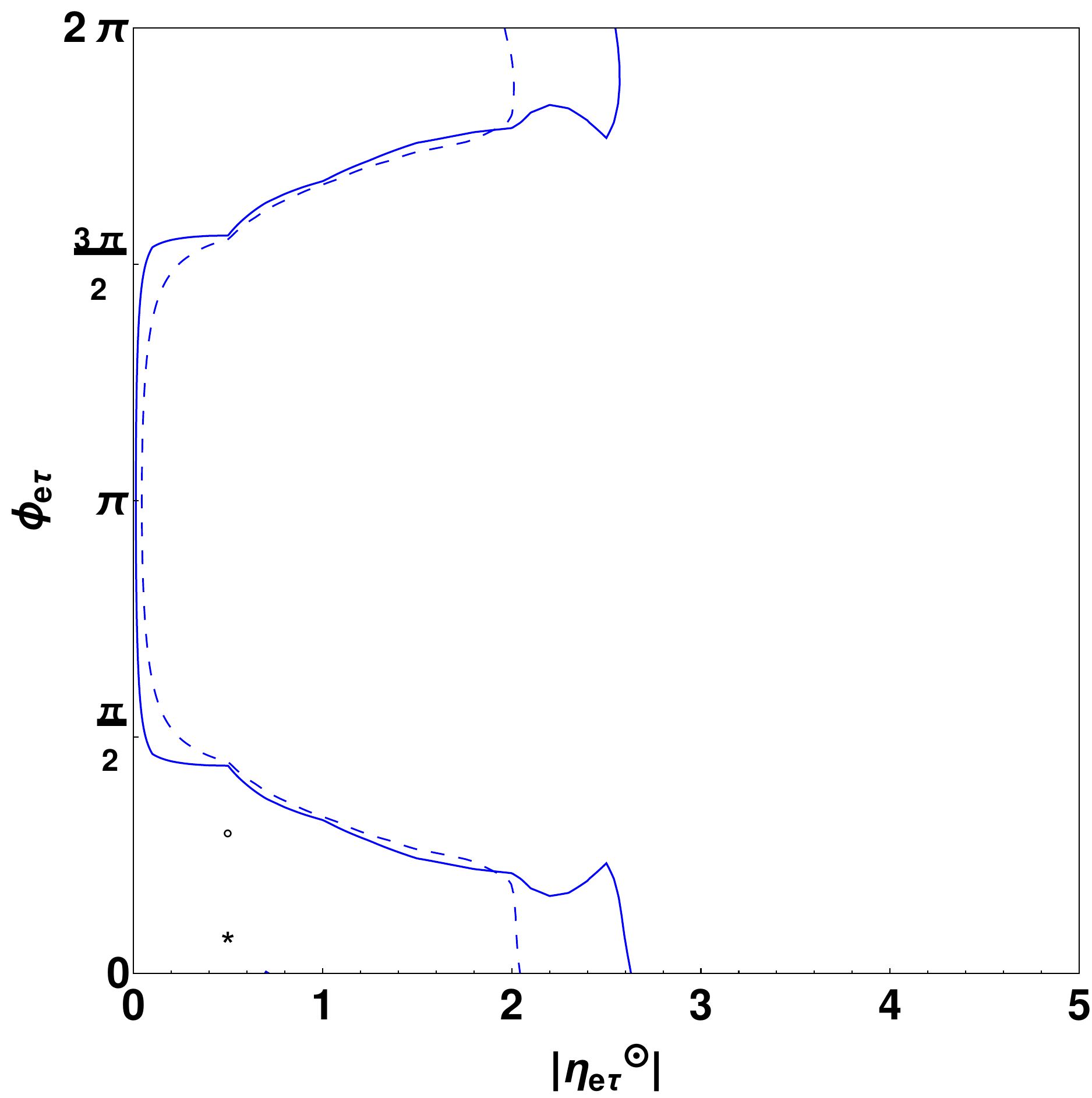}
    \includegraphics[width=4.5cm,height=4.5cm]{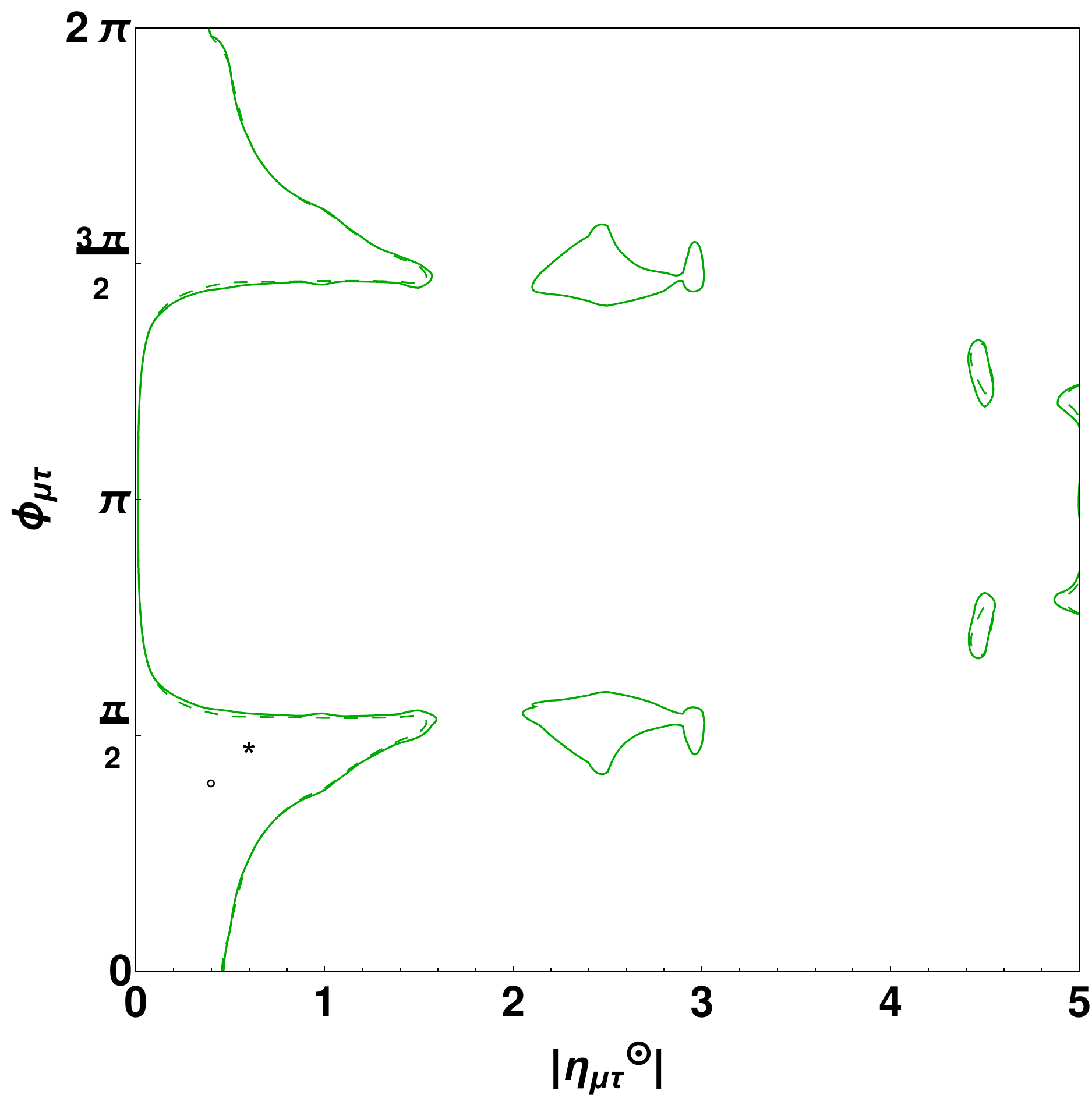}
    \caption{Same as Fig.~\ref{fig:nondiagonalchi} including the KamLAND (solid) and JUNO (dashed) prior on the solar parameters.}
    \label{fig:nondiagonalchi_phi_Kamland}
\end{figure}

The allowed regions in the parameters space of the NSI parameters ($\eta_{\alpha\alpha}^\odot-m_1$ for diagonal parameters and $|\eta_{\alpha\beta}^\odot|-\phi_{\alpha\beta}$ for the non diagonal parameters) are shown in Fig.~\ref{fig:diagonalchi_m1_Kamland} and \ref{fig:nondiagonalchi_phi_Kamland} where solid lines are for KamLAND and dashed ones for JUNO. The inclusion of the KamLAND prior reduces the allowed values of the diagonal parameters as expected.
Moreover, the preferred values of the lightest of the neutrino masses is lowered; in the case of $\eta_{\mu\mu}^\odot$, the fit prefers $m_1<10^{-4}$ eV. For the off-diagonal parameters, we observe the same behavior, with $|\eta_{e\mu}^\odot|$ which can assume large values only if $\phi_{e\mu}$ is around $\pi$, $|\eta_{e\tau}^\odot|$ if $\phi_{e\tau}$ is around 0 and $|\eta_{\mu\tau}^\odot|$ only for $\phi_{\mu\tau}\sim\pm\pi/2$.

Regarding JUNO, we expect that the prior of the mass splitting will only play a minor role with respect to the KamLAND case. Indeed, given that solar data can only weakly constrain the mass splitting, both priors have the effect of fixing the mass splitting in our analyses. On the other hand, the strong bound on the mixing angle can have a strong impact on the allowed regions, reducing them in the case of $\eta_{\mu\mu}^\odot$, $\eta_{e\mu}^\odot$ and $\eta_{\mu\tau}^\odot$. In the case of $\eta_{ee}^\odot$, $\eta_{\tau\tau}^\odot$ and $\eta_{e\tau}^\odot$, on the other hand, we observe a more pronounced shape change in the allowed regions. This is made possible by the fact that the preferred value of $\theta_{12}$ using solar data is larger than the one used to add the JUNO prior. This resulted in a slightly increased minimum $\chi^2$ at the $\lesssim2\sigma$ level. 

\begin{figure}
\centering
\includegraphics[width=0.32\textwidth]{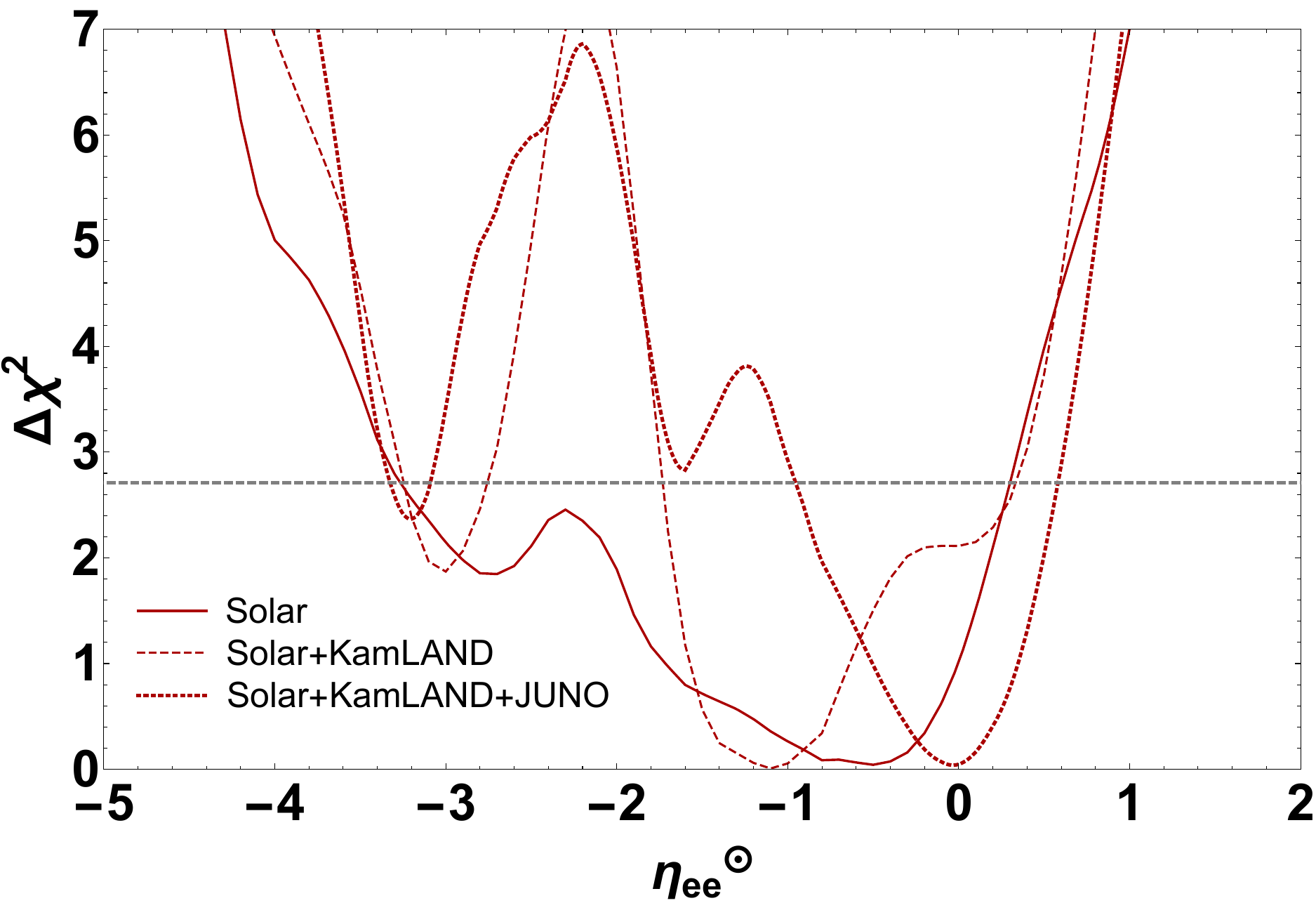}
\includegraphics[width=0.32\textwidth]{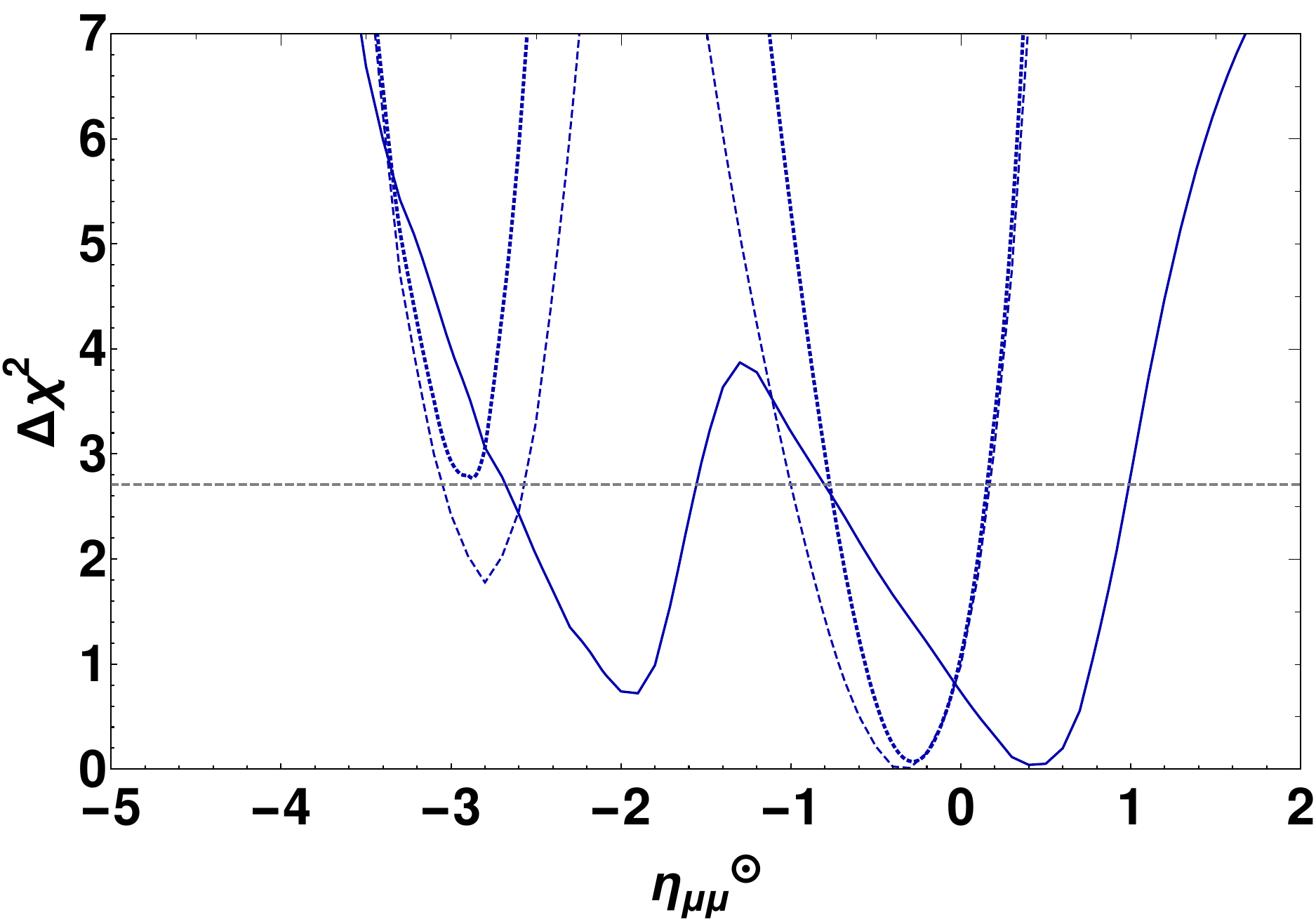}
\includegraphics[width=0.32\textwidth]{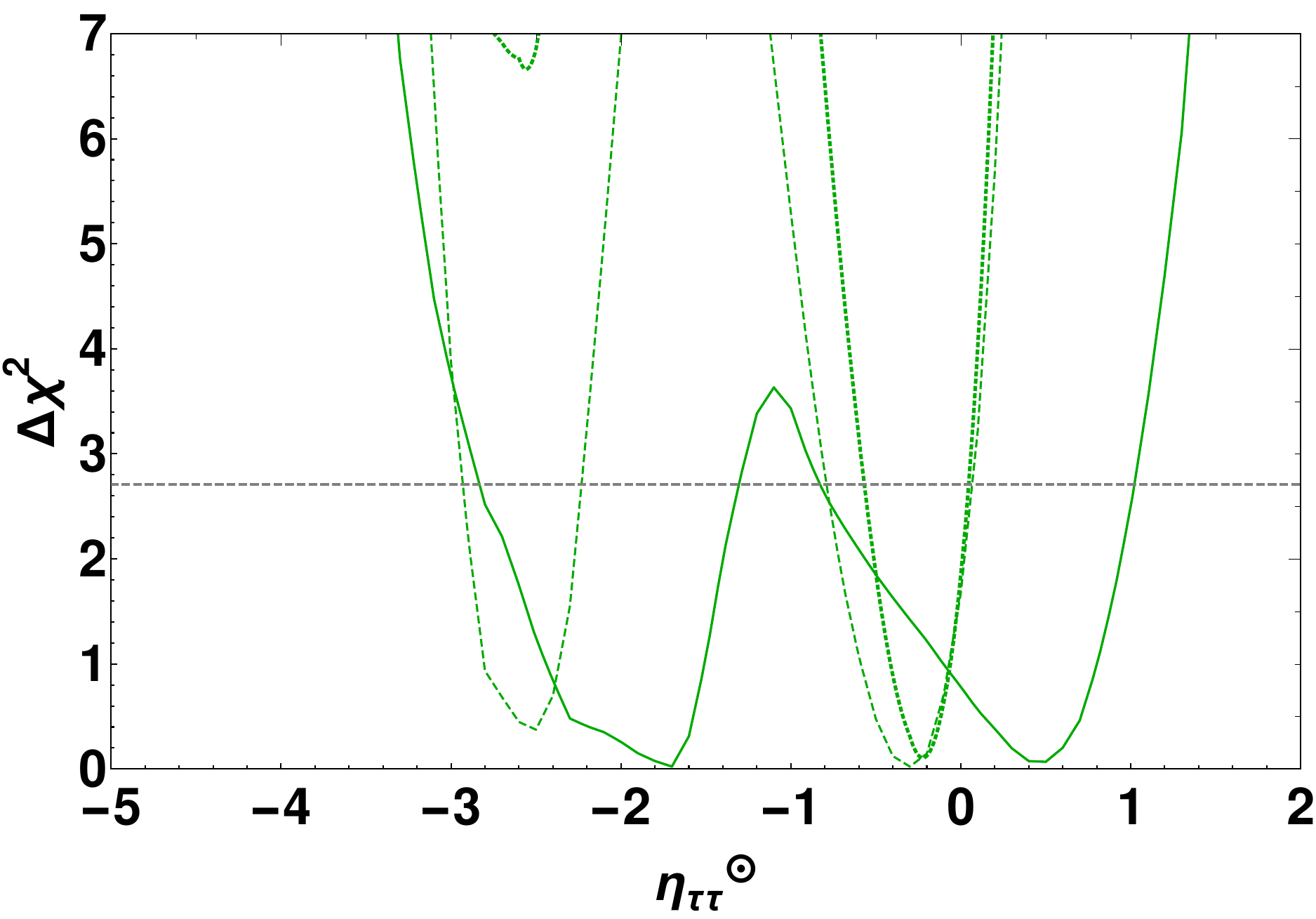}
\includegraphics[width=0.32\textwidth]{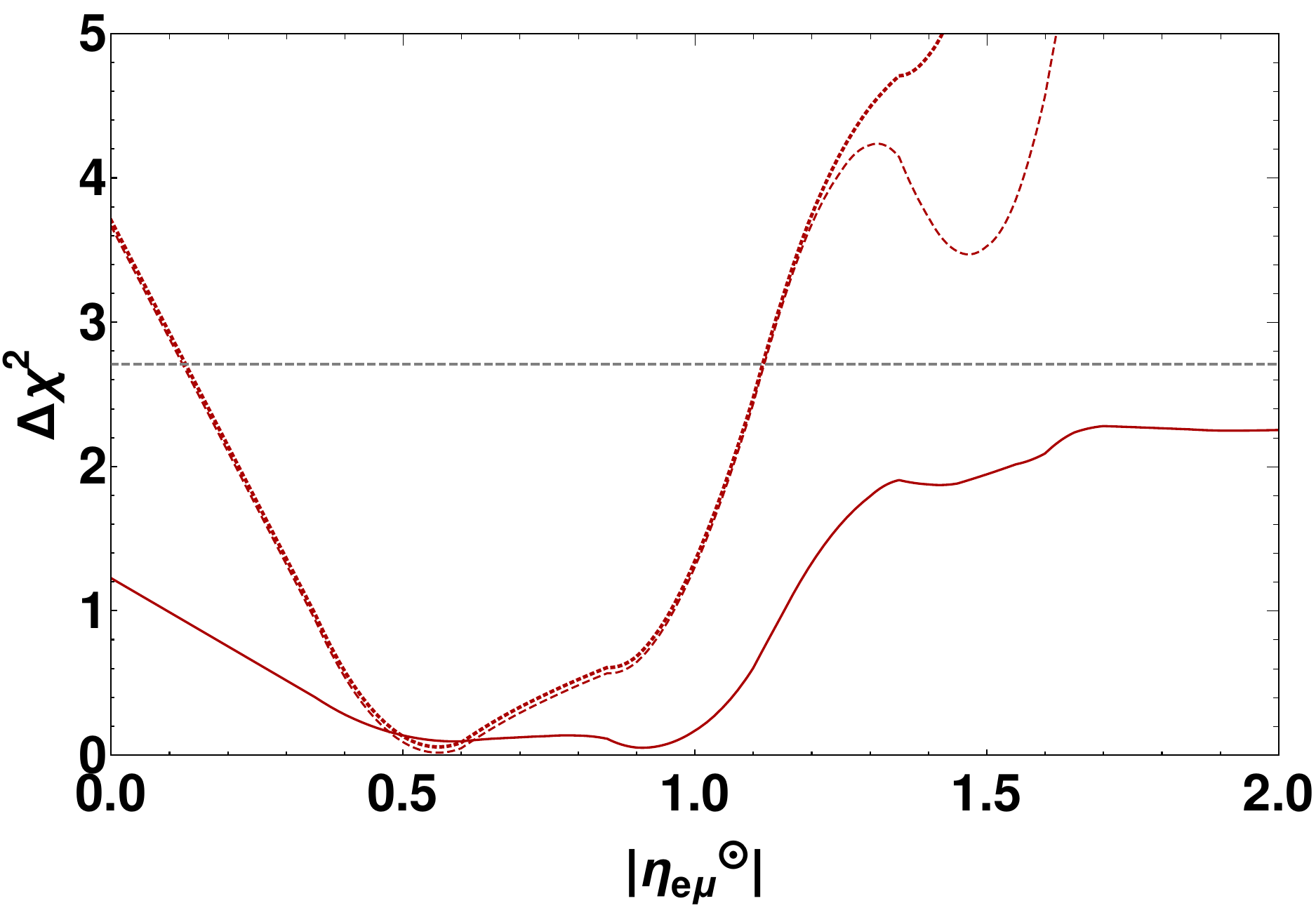}
\includegraphics[width=0.32\textwidth]{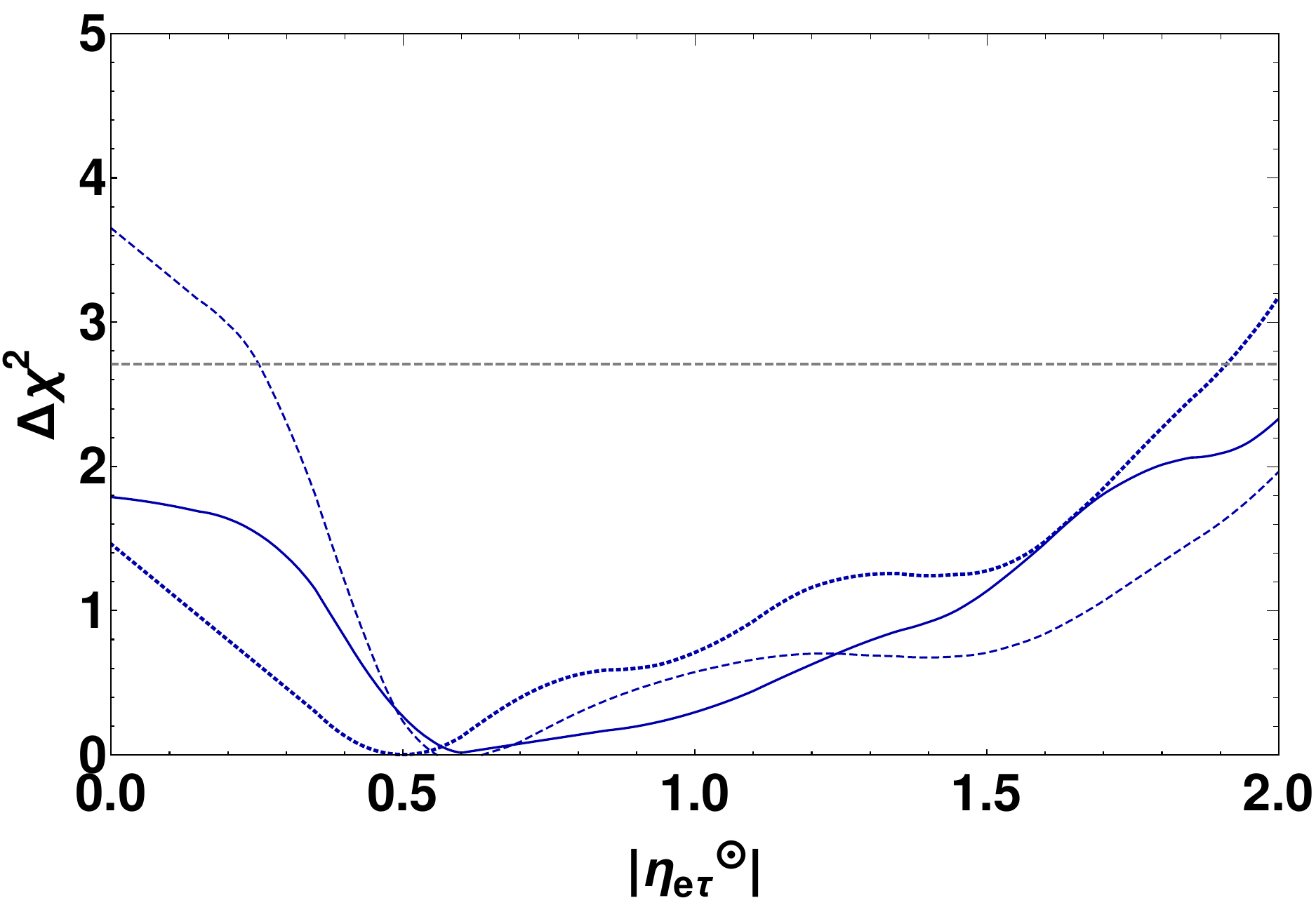}
\includegraphics[width=0.32\textwidth]{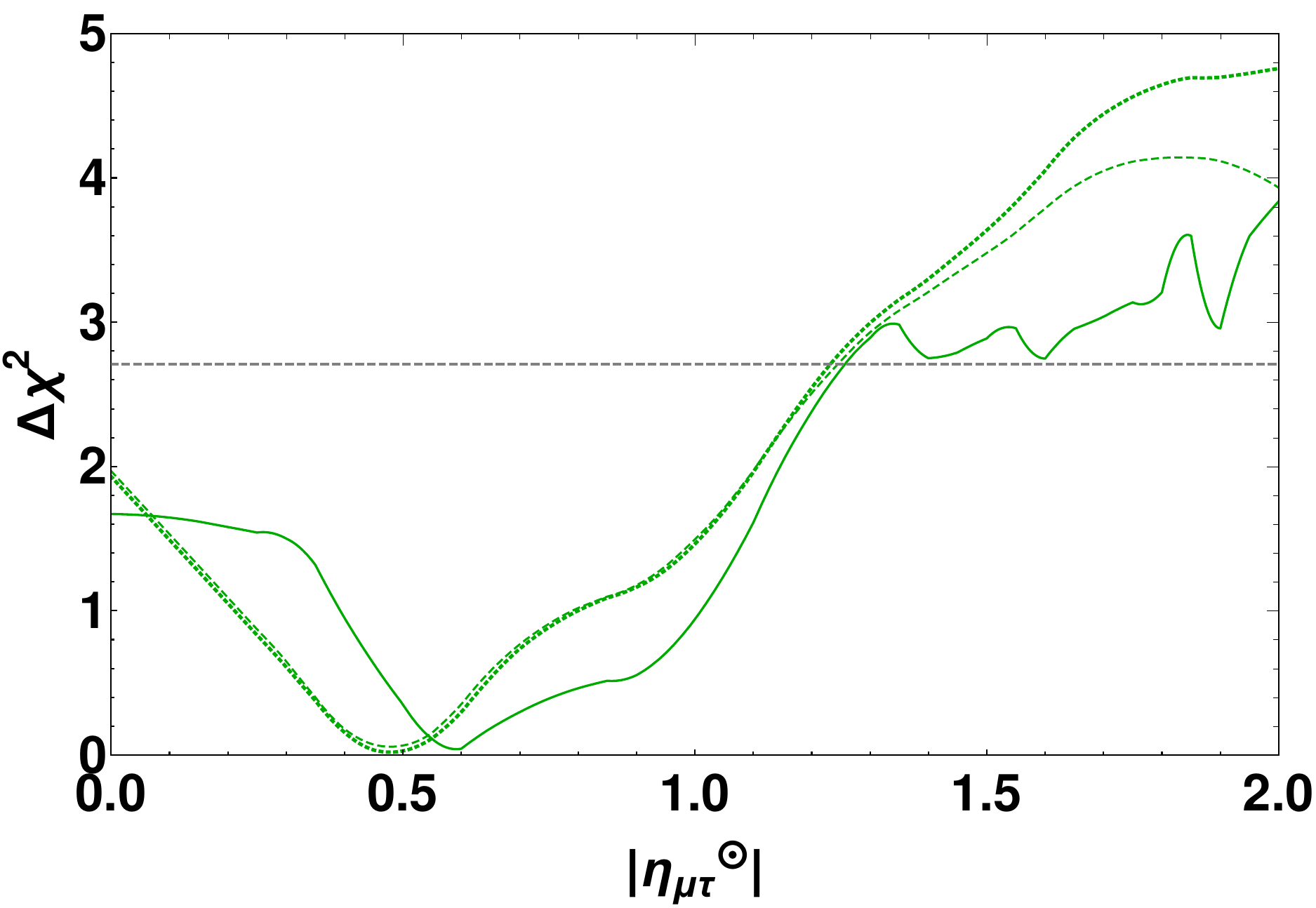}
\caption{The 1D $\Delta\chi^2$'s for the three diagonal parameters (top row) and the absolute value of the three off-diagonal parameters (bottom row).
The solid lines are solar data only and the dashed lines are solar data and KamLAND.
The dotted lines are the expected sensitivity including solar data, KamLAND, and JUNO.
A horizontal line is drawn at the 90\% CL point.}
\label{fig:1D}
\end{figure}

All of these results are summarized in table \ref{tab:summary}, where we show the 90\% CL limits on the parameters (1 degree of freedom) and fig.~\ref{fig:1D}. With $\Re\eta_{\alpha\beta}^\odot$ ($\Im\eta_{\alpha\beta}^\odot$) we refer to the limits on off-diagonal scalar NSI parameters obtained for $\phi_{\alpha\beta}=0,\pi$ ($\phi_{\alpha\beta}=\pm\pi/2$).
We note that for comparison the Borexino only results from \cite{Ge:2018uhz} translated to our units at 90\% are approximately $-38\le\eta_{ee}^\odot\le13$ and $-18\le\eta_{\mu\mu,\tau\tau}^\odot\le27$ with a low significance preferred value of $\eta_{ee}^\odot=-36$.
Thus our results with Solar and KamLAND data represent an improvement over existing constraints by about 1 to 1.5 orders of magnitude with the possibility for additional improvement with JUNO's reactor measurement of the solar parameters.
In the units of \cite{Babu:2019iml}, we find that the upper limit on $\Delta m_{\rm Sun}\lesssim0.8$ meV depending on the neutrino flavor and matter fermion, also an order of magnitude improvement over existing constraints.
We note that, as in the case of \cite{Babu:2019iml}, the constraint on the couplings $y_ay_{\alpha\beta}$ will weaken for mediators lighter than $m_\phi\sim10^{-14}$ eV.

\begin{table}
\centering
\caption{The 90\% allowed regions (for 1 degree of freedom) on the various sNSI parameters with different experimental inputs at $m_1=0$.
The Solar+JUNO column is the expected future sensitivity.}
\label{tab:summary}
\begin{tabular}{c|c|c|c}
 & Solar & Solar+KamLAND & Solar+JUNO \\\hline
$\eta_{ee}^\odot$ & [-3.21,0.27] & [-1.22,0.26]  & [-0.97,0.59] \\
$\eta_{\mu\mu}^\odot$  & [-2.58,-1.60]$\oplus$[-0.70,0.94] & [-2.87,-2.50]$\oplus$[-1.01,0.15] & [-0.79,0.11] \\
$\eta_{\tau\tau}^\odot$ & [-2.57,-1.35]$\oplus$[-0.72,1.02] & [-2.57,-2.28]$\oplus$[-0.74,0.06] & [-0.56,0.05] \\\hline
$\Re\eta_{e\mu}^\odot$ & [-2.53,1.57] & [-1.07,-0.13] & [-1.03,-0.15] \\
$\Im\eta_{e\mu}^\odot$ & [-1.51,1.49] & $<$0.01 & $<$0.01 \\
$\Re\eta_{e\tau}^\odot$ & [2.05,0.05] & [0.42,2.05] & [0.0,0.13] $\oplus$ [0.95,1.56] \\
$\Im\eta_{e\tau}^\odot$ & [-0.61,0.61] & [-0.02,0.02] & [-0.02,0.02] \\
$\Re\eta_{\mu\tau}^\odot$ & [0.0,0.04] & [0.0,0.03] & [0.0,0.01] \\
$\Im\eta_{\mu\tau}^\odot$ & [-0.55,0.55] & [-1.1,1.1] & [-1.0,1.0] 
\end{tabular}
\end{table}

\section{Conclusions}
Scalar non-standard neutrino interactions (sNSI) is a compelling new physics scenario where a new interaction creates a medium dependent contribution to the mass of a neutrino.
This modifies the propagation of neutrinos in a non-trivial way. 
Given the general concordance that is emerging in the standard three-flavor oscillation picture -- consistent measurements of $\Delta m^2_{21}$ from solar and reactor experiments and consistent measurements of $\Delta m^2_{31}$ in reactor, accelerator, and atmospheric experiments -- it is possible to do a complete fit to relevant oscillation data to constrain new physics scenarios such as sNSI.

Unlike the more often studied vector NSI whose effect increases with energy as well as density, the sNSI effect only increases with density, thus different experiments will be optimal in different cases.
It is for this reason that we have focused on solar neutrinos and provided priors from terrestrial experiments in low-density experiments which are expected to be about 1-2 orders of magnitude less sensitive.

For the solar neutrino analysis we included data from Borexino and SNO and then set the oscillation parameters based on data from KamLAND.
We also estimated the future sensitivity to sNSI with the improved measurements of the regular oscillation parameters with JUNO.

We found some constraints on all parameters from solar data alone.
The addition of KamLAND data improves them considerably and our constraints are 1-1.5 orders of magnitude stronger than those existing in the literature.
We also see that improving measurements even in relatively low density environments via JUNO will further improve our constraints on sNSI parameters.

Finally, we point out the crucial fact that sNSI is fundamentally intertwined with the absolute neutrino mass scale and a possible detection of sNSI would also provide new information about the mass scale that is very complementary to other searches from cosmology or end point searches.

\acknowledgments
We thank Evgeny Akhmedov for helpful comments.
PBD acknowledges support from the US Department of Energy under Grant Contract DE-SC0012704.

\bibliographystyle{JHEP}
\bibliography{SolarNSI}

\end{document}